%% file: DESY-03-201.tex
\documentclass[znote,zbstdefault]{./LaTeX/zeus/zeus_paper}
\usepackage[english]{babel}
\input ./LaTeX/zeus/zeus_def.tex
\input ./LaTeX/user/def.tex
\input  InstPaper-cit.tex

\begin{document}
\include{InstPaper-tit}
\include{auth113_out}

\include{InstPaper-txt}

\include{InstPaper-ref}

\include{InstPaper-tab}
\include{InstPaper-fig}

%
%
\end{document}

%% file: LaTeX/zeus/zeus_def.tex
\newcommand{\ZcoosysA}{%
The ZEUS coordinate system is a right-handed Cartesian system, with the $Z$
axis pointing in the proton beam direction, referred to as the ``forward
direction'', and the $X$ axis pointing left towards the center of HERA.
The coordinate origin is at the nominal interaction point.\xspace}
\newcommand{\ZcoosysB}{%
The ZEUS coordinate system is a right-handed Cartesian system, with the $Z$
axis pointing in the proton beam direction, referred to as the ``forward
direction'', and the $X$ axis pointing left towards the centre of HERA.
The coordinate origin is at the nominal interaction point.\xspace}
\newcommand{\Zpsrap}{%
The pseudorapidity is defined as $\eta=-\ln\left(\tan\frac{\theta}{2}\right)$,
where the polar angle, $\theta$, is measured with respect to the proton beam
direction.\xspace}
\newcommand{\ZcoosysfnA}{\footnote{\ZcoosysA}}
\newcommand{\ZcoosysfnB}{\footnote{\ZcoosysB}}
\newcommand{\ZcoosysfnAeta}{\footnote{\ZcoosysA\Zpsrap}}
\newcommand{\ZcoosysfnBeta}{\footnote{\ZcoosysB\Zpsrap}}
\newcommand{\Zdetdesc}{%
A detailed description of the ZEUS detector can be found 
elsewhere~\cite{zeus:1993:bluebook}. A brief outline of the 
components that are most relevant for this analysis is given
below.\xspace}
\newcommand{\Zctddesc}[1]{%
Charged particles are tracked in the central tracking detector (CTD)~\citeCTD,
which operates in a magnetic field of $1.43\Tesla$ provided by a thin 
superconducting coil. The CTD consists of 72~cylindrical drift chamber 
layers, organized in 9~superlayers covering the polar-angle#1 region 
\mbox{$15^\circ<\theta<164^\circ$}. The transverse-momentum resolution for
full-length tracks is $\sigma(p_T)/p_T=0.0058p_T\oplus0.0065\oplus0.0014/p_T$,
with $p_T$ in $\Gev$.}
\newcommand{\Zcaldesc}{%
The high-resolution uranium--scintillator calorimeter (CAL)~\citeCAL consists 
of three parts: the forward (FCAL), the barrel (BCAL) and the rear (RCAL)
calorimeters. Each part is subdivided transversely into towers and
longitudinally into one electromagnetic section (EMC) and either one (in RCAL)
or two (in BCAL and FCAL) hadronic sections (HAC). The smallest subdivision of
the calorimeter is called a cell.  The CAL energy resolutions, as measured under
test-beam conditions, are $\sigma(E)/E=0.18/\sqrt{E}$ for electrons and
$\sigma(E)/E=0.35/\sqrt{E}$ for hadrons ($E$ in $\Gev$).}
\newcommand{\xxx}{\hbox{$\qsm\qsm\qsm$}\xspace}
\newcommand{\DO}{{D{\O}}\xspace}
\newcommand{\Anti}[1]{\hbox{(anti-)}\discretionary{}{}{}#1}
\newcommand{\anti}[1]{(anti\discretionary{-)}{}{)}#1}
\chardef\usc=95
\chardef\til=126
\catcode`\@=11 
\DeclareRobustCommand\xdotspace{\futurelet\@let@token\@xdotspace}
\def\@xdotspace{%
  \ifx\@let@token.\else
  \ifx\@let@token\bgroup.\else
  \ifx\@let@token\egroup.\else
  \ifx\@let@token\/.\else
  \ifx\@let@token\ .\else
  \ifx\@let@token~.\else
  \ifx\@let@token!.\else
  \ifx\@let@token,.\else
  \ifx\@let@token:.\else
  \ifx\@let@token;.\else
  \ifx\@let@token?.\else
  \ifx\@let@token/.\else
  \ifx\@let@token'.\else
  \ifx\@let@token).\else
  \ifx\@let@token-.\else
  \ifx\@let@token\@xobeysp.\else
  \ifx\@let@token\space.\else
  \ifx\@let@token\@sptoken.\else
   .\space
   \fi\fi\fi\fi\fi\fi\fi\fi\fi\fi\fi\fi\fi\fi\fi\fi\fi\fi}
\catcode`\@=12 
\newcommand{\CL}[1]{$#1\%$~C.L\xdotspace}
\newcommand{\stru}[2]{%
   \relax\ifmmode\hbox{\vrule height#1 depth#2 width0pt}%
   \else\vrule height#1 depth#2 width0pt\fi}
\newcommand{\uline}[1]{$\underline{\hbox{#1\stru{0.pt}{.485ex}}}$}
\newcommand{\Ronum}[1]{\uppercase\expandafter{\romannumeral#1}}
\newcommand{\ronum}[1]{\expandafter{\romannumeral#1}}
\DeclareRobustCommand{\LaTeXZ}{%
  \LaTeX\kern-.05em4\kern-.1em
  {\raisebox{-0.2ex}{$\scriptstyle\text{ZEUS}$}}\xspace}
\newcommand{\Auml}{\"A}
\newcommand{\Ouml}{\"O}
\newcommand{\Uuml}{\"U}
\newcommand{\auml}{\"a}
\newcommand{\ouml}{\"o}
\newcommand{\uuml}{\"u}
\newcommand{\eq}[1]{(\ref{eq-#1})}
\newcommand{\eqsand}[2]{(\ref{eq-#1}) and~(\ref{eq-#2})}
\newcommand{\eqsto}[2]{(\ref{eq-#1})--(\ref{eq-#2})}
\newcommand{\eqstwo}[2]{(\ref{eq-#1},\ref{eq-#2})}
\newcommand{\eqsthr}[3]{(\ref{eq-#1},\ref{eq-#2},\ref{eq-#3})}
\newcommand{\eqsfou}[4]{(\ref{eq-#1},\ref{eq-#2},\ref{eq-#3},\ref{eq-#4})}
\newcommand{\Eq}[1]{Equation~(\ref{eq-#1})}
\newcommand{\Eqsto}[2]{Equations~(\ref{eq-#1})--(\ref{eq-#2})}
\newcommand{\fig}[1]{Fig.~\ref{fig-#1}}
\newcommand{\Fig}[1]{Figure~\ref{fig-#1}}
\newcommand{\figand}[2]{Figs.~\ref{fig-#1} and~\ref{fig-#2}}
\newcommand{\Figand}[2]{Figures~\ref{fig-#1} and~\ref{fig-#2}}
\newcommand{\tab}[1]{Table~\ref{tab-#1}}
\newcommand{\Tab}[1]{Table~\ref{tab-#1}}
\newcommand{\taband}[2]{Tables~\ref{tab-#1} and~\ref{tab-#2}}
\newcommand{\cha}[1]{Chap.~\ref{cha-#1}}
\newcommand{\Cha}[1]{Chapter~\ref{cha-#1}}
\newcommand{\sect}[1]{Sect.~\ref{sec-#1}}
\newcommand{\Sect}[1]{Section~\ref{sec-#1}}
\newcommand{\sectand}[2]{Sects.~\ref{sec-#1} and~\ref{sec-#2}}
\newcommand{\sectto}[2]{Sects.~\ref{sec-#1} to~\ref{sec-#2}}
\newcommand{\Sectand}[2]{Sections~\ref{sec-#1} and~\ref{sec-#2}}
\newcommand{\Sectto}[2]{Sections~\ref{sec-#1} to~\ref{sec-#2}}
\newcommand{\LogMess}[1]{\typeout{^^J==> #1^^J}}
\DeclareMathAlphabet{\mathbf}{OT1}{cmr}{bx}{sl}
\newcommand{\eVdist}{\kern-0.06667em}
\newcommand{\Ev}{{\text{e}\eVdist\text{V\/}}}
\newcommand{\Kev}{{\text{ke}\eVdist\text{V\/}}}
\newcommand{\Mev}{{\text{Me}\eVdist\text{V\/}}}
\newcommand{\Gev}{{\text{Ge}\eVdist\text{V\/}}}
\newcommand{\Tev}{{\text{Te}\eVdist\text{V\/}}}
\newcommand{\ev}{{\,\text{e}\eVdist\text{V\/}}}
\newcommand{\kev}{{\,\text{ke}\eVdist\text{V\/}}}
\newcommand{\mev}{{\,\text{Me}\eVdist\text{V\/}}}
\newcommand{\gev}{{\,\text{Ge}\eVdist\text{V\/}}}
\newcommand{\tev}{{\,\text{Te}\eVdist\text{V\/}}}
\newcommand{\mb}{\,\text{mb}}
\newcommand{\mub}{\,\upmu\text{b}}
\newcommand{\nb}{\,\text{nb}}
\newcommand{\pb}{\,\text{pb}}
\newcommand{\fb}{\,\text{fb}}
\newcommand{\mubi}{\,\upmu\text{b}^{-1}}
\newcommand{\mbi}{\,\text{mb}^{-1}}
\newcommand{\nbi}{\,\text{nb}^{-1}}
\newcommand{\pbi}{\,\text{pb}^{-1}}
\newcommand{\fbi}{\,\text{fb}^{-1}}
\newcommand{\met}{\,\text{m}}
\newcommand{\nm}{\,\text{nm}}
\newcommand{\mum}{\,\upmu\text{m}}
\newcommand{\mm}{\,\text{mm}}
\newcommand{\cm}{\,\text{cm}}
\newcommand{\km}{\,\text{km}}
\newcommand{\Amp}{\,\text{A}}
\newcommand{\nA}{\,\text{nA}}
\newcommand{\muA}{\,\upmu\text{A}}
\newcommand{\mA}{\,\text{mA}}
\newcommand{\kA}{\,\text{kA}}
\newcommand{\Hz}{\,\text{Hz}}
\newcommand{\kHz}{\,\text{kHz}}
\newcommand{\MHz}{\,\text{MHz}}
\newcommand{\GHz}{\,\text{GHz}}
\newcommand{\ppm}{\,\text{ppm}}
\newcommand{\ppb}{\,\text{ppb}}
\newcommand{\scd}{\,\text{s}}
\newcommand{\ps}{\,\text{ps}}
\newcommand{\ns}{\,\text{ns}}
\newcommand{\mus}{\,\upmu\text{s}}
\newcommand{\ms}{\,\text{ns}}
\newcommand{\rad}{\,\text{rad}}
\newcommand{\murad}{\,\upmu\text{rad}}
\newcommand{\mrad}{\,\text{mrad}}
\newcommand{\gra}{\,\text{g}}
\newcommand{\pg}{\,\text{pg}}
\newcommand{\nang}{\,\text{ng}}
\newcommand{\mug}{\,\upmu\text{g}}
\newcommand{\mg}{\,\text{mg}}
\newcommand{\kg}{\,\text{kg}}
\newcommand{\Tesla}{\,\text{T}}
\newcommand{\Kelvin}{\,\text{K}}
\newcommand{\odiv}{{\operatorname*{div}}}
\newcommand{\ograd}{{\operatorname*{grad}}}
\newcommand{\orot}{{\operatorname*{rot}}}
\newcommand{\odiag}{{\operatorname*{diag}}}
\newcommand{\ocov}{{\operatorname*{cov}}}
\newcommand{\slashfrac}[2]{%
  \raisebox{0.5ex}{\ensuremath #1}\kern-0.12em/\kern-0.08em
  \raisebox{-.8ex}{\ensuremath #2}}
\newcommand{\intl}{\int\limits}
\newcommand{\ointl}{\oint\limits}
\newcommand{\sqr}[3]{%
    {\vcenter{\hrule height.#3ex\hbox{\vrule width.#2ex height#1ex
     \kern#1ex\vrule width.#3ex}\hrule height.#2ex}}}
\newcommand{\dalem}{\mathchoice\sqr{1.3}{08}{24}\sqr{1.4}{08}{24}
                    \sqr{1.15}{07}{21}\sqr{1.0}{06}{18}\,}
\newcommand{\vect}[1]{\begin{matrix}#1\end{matrix}}
\newcommand{\pvect}[1]{\begin{pmatrix}#1\end{pmatrix}}
\newcommand{\bvect}[1]{\begin{bmatrix}#1\end{bmatrix}}
\newcommand{\vvect}[1]{\begin{vmatrix}#1\end{vmatrix}}
\newcommand{\Vvect}[1]{\begin{Vmatrix}#1\end{Vmatrix}}
\newcommand{\widebar}[1]{%
   \mkern1.5mu\overline{\mkern-1.5mu#1\mkern-1.mu}\mkern1.mu}
\catcode`\@=11 
\newcommand{\parenbar}{\mathpalette\p@renb@r}
\def\p@renb@r#1#2{\vbox{%
  \ifx#1\scriptscriptstyle \dimen@.7em\dimen@ii.2em\else
  \ifx#1\scriptstyle \dimen@.8em\dimen@ii.25em\else
  \dimen@1em\dimen@ii.4em\fi\fi \offinterlineskip
  \ialign{\hfill##\hfill\cr
    \vbox{\hrule width\dimen@ii}\cr
    \noalign{\vskip-.3ex}%
    \hbox to\dimen@{$\mathchar300\hfil\mathchar301$}\cr
    \noalign{\vskip-.3ex}%
    $#1#2$\cr}}}
\catcode`\@=12 
\newcommand{\nuan}{\parenbar{\nu}}
\newcommand{\nunubar}{\parenbar{\nu}}
\newcommand{\ppbar}{\parenbar{p}}
\newcommand{\qqbar}{\parenbar{q}}
\newcommand{\pbar}{\widebar{p}}
\newcommand{\nbar}{\widebar{n}}
\newcommand{\qbar}{\widebar{q}}
\newcommand{\dbar}{\widebar{d}}
\newcommand{\ubar}{\widebar{u}}
\newcommand{\sbar}{\widebar{s}}
\newcommand{\cbar}{\widebar{c}}
\newcommand{\bbar}{\widebar{b}}
\newcommand{\tbar}{\widebar{t}}
\newcommand{\nubar}{\widebar{\nu}}
\newcommand{\Dbar}{\widebar{D}}
\newcommand{\ebar}{\widebar{e}}
\newcommand{\Ebar}{\widebar{e}}
\newcommand{\Hebar}{\widebar{\text{He}}}
\newcommand{\Cbar}{\widebar{\text{C}}}
\newcommand{\Abar}{\widebar{\text{A}}}
\newcommand{\Kbar}{\widebar{K}}
\newcommand{\chibar}{\widebar{\chi}}
\newcommand{\LQbar}{\widebar{\rm LQ}}
\newcommand{\zero}{{\scriptscriptstyle 0}}
\newcommand{\sone}{{\scriptscriptstyle 1}}
\newcommand{\stwo}{{\scriptscriptstyle 2}}
\newcommand{\sthr}{{\scriptscriptstyle 3}}
\newcommand{\sfou}{{\scriptscriptstyle 4}}
\newcommand{\sfiv}{{\scriptscriptstyle 5}}
\newcommand{\ssix}{{\scriptscriptstyle 6}}
\newcommand{\ssev}{{\scriptscriptstyle 7}}
\newcommand{\seig}{{\scriptscriptstyle 8}}
\newcommand{\snin}{{\scriptscriptstyle 9}}
\newcommand{\sten}{{\scriptscriptstyle 10}}
\newcommand{\half}{{\scriptscriptstyle 1/2}}
\newcommand{\MSbar}{\hbox{$\overline{\rm MS}$}\xspace}
\newcommand{\UO}{{\rm U}(1)_\sY}
\newcommand{\SUT}{{\rm SU}(2)_\sL}
\newcommand{\SUTH}{{\rm SU}(3)_c}
\newcommand{\SUTUO}{{\rm SU}(2)_\sL\times{\rm U}(1)_\sY}
\newcommand{\SUTSUTUO}{{\rm SU}(3)_c\times{\rm SU}(2)_\sL\times{\rm U}(1)_\sY}
\newcommand{\alsmu}[1]{{\alpha_s(\mu_{#1}^2)}}
\newcommand{\alsmz}{{\alpha_s(M_Z^2)}}
\newcommand{\alsqs}{{\alpha_s(Q^2)}}
\newcommand{\als}{\alpha_s}
\newcommand{\ctwb}{\cos\theta_W}
\newcommand{\ctws}{\cos^2\theta_W}
\newcommand{\diff}{{\rm d}}
\newcommand{\gh}{{\gamma_h}}
\newcommand{\lsq}[1]{{\lambda_{#1}}}
\newcommand{\lsqp}[1]{{\lambda'_{#1}}}
\newcommand{\lsqpp}[1]{{{\lambda''}_{#1}}}
\newcommand{\rnge}{\hbox{$\,\text{--}\,$}}
\newcommand{\sihat}{{\hat\sigma}}
\newcommand{\sitil}{{\tilde\sigma}}
\newcommand{\stwb}{\sin\theta_W}
\newcommand{\stwss}{\sin^4\theta_W}
\newcommand{\stws}{\sin^2\theta_W}
\newcommand{\ta}{{\theta^\ast}}
\newcommand{\tilR}{\tilde R}
\newcommand{\tilS}{\tilde S}
\newcommand{\tilU}{\tilde U}
\newcommand{\tilV}{\tilde V}
\newcommand{\tl}{{\theta_\ell}}
\newcommand{\tw}{\theta_W}
\newcommand{\Born}{{\rm Born}}
\newcommand{\BR}{{\rm BR}}
\newcommand{\CC}{{\rm CC}}
\newcommand{\CI}{{\rm CI}}
\newcommand{\CJ}{{\rm CJ}}
\newcommand{\DA}{{\rm DA}}
\newcommand{\DCA}{{\rm DCA}}
\newcommand{\GRV}{{\rm GRV}}
\newcommand{\IC}{{\rm IC}}
\newcommand{\ISR}{{\rm ISR}}
\newcommand{\IP}{{\rm I$\kern-0.01667em$P}\xspace}
\newcommand{\JB}{{\rm JB}}
\newcommand{\LL}{{\rm LL}}
\newcommand{\LQ}{{\rm LQ}}
\newcommand{\LR}{{\rm LR}}
\newcommand{\MC}{{\rm MC}}
\newcommand{\NC}{{\rm NC}}
\newcommand{\NS}{{\rm NS}}
\newcommand{\QPM}{{\rm QPM}}
\newcommand{\RL}{{\rm RL}}
\newcommand{\RPv}{{{\not R}_P}}
\newcommand{\RR}{{\rm RR}}
\newcommand{\SM}{{\rm SM}}
\newcommand{\bkg}{{\rm bkg}}
\newcommand{\cut}{{\rm cut}}
\newcommand{\data}{{\rm data}}
\newcommand{\sdet}{{\rm det}}
\newcommand{\dof}{{\rm dof}}
\newcommand{\eff}{{\rm eff}}
\newcommand{\elm}{{\rm elm}}
\newcommand{\evt}{{\rm evt}}
\newcommand{\exc}{{\rm exc}}
\newcommand{\fit}{{\rm fit}}
\newcommand{\had}{{\rm had}}
\newcommand{\jet}{{\rm jet}}
\newcommand{\meas}{{\rm meas}}
\newcommand{\miss}{{\rm miss}}
\newcommand{\obs}{{\rm obs}}
\newcommand{\pnt}{{\rm pnt}}
\newcommand{\sL}{{\rm L}}
\newcommand{\sR}{{\rm R}}
\newcommand{\sS}{{\rm S}}
\newcommand{\sT}{{\rm T}}
\newcommand{\sY}{{\rm Y}}
\newcommand{\sca}{{\rm sca}}
\newcommand{\sexp}{{\rm exp}}
\newcommand{\sint}{{\rm int}}
\newcommand{\smin}{{\rm min}}
\newcommand{\smax}{{\rm max}}
\newcommand{\srad}{{\rm rad}}
\newcommand{\sthe}{{\rm the}}
\newcommand{\sys}{{\rm sys}}
\newcommand{\tot}{{\rm tot}}
\newcommand{\true}{{\rm true}}
\newcommand{\Cor}{{\cal C}}
\newcommand{\F}{{\cal F}}
\newcommand{\Lumi}{{\cal L}}
\newcommand{\Prob}{{\cal P}}
\newcommand{\ord}[1]{{\cal O}(#1)}
\newcommand{\ordlr}[1]{{\cal O}\left(#1\right)}
\mathchardef\qsm=63
\mathchardef\pls=43
\mathchardef\mns=512
\mathchardef\plm=518
\mathchardef\eql=61
\mathchardef\smallleft=300
\mathchardef\smallright=301
\mathchardef\les=316
\mathchardef\gre=318
\mathchardef\leq=532
\mathchardef\grq=533
\newcommand{\cA}{{\phantom{0}}}
\newcommand{\cB}{{\phantom{00}}}
\newcommand{\cC}{{\phantom{000}}}
\newcommand{\cD}{{\phantom{0000}}}
\newcommand{\cAp}{{\phantom{0.}}}
\newcommand{\cM}{{\phantom{\mns}}}
\newcommand{\cP}{{\phantom{\pls}}}
\catcode`\@=11 
\newcounter{pict@width}
\newcounter{pict@height}
\newlength{\pict@scale}
\newcommand{\psfigadd}[4]{%
\setcounter{pict@width}{1*\ratio{#2+\pict@scale/2}{\pict@scale}}
\setcounter{pict@height}{1*\ratio{#3+\pict@scale/2}{\pict@scale}}
\setlength{\unitlength}{\pict@scale}
\hbox to #2{\hspace{-\fill}\begin{picture}(\thepict@width,\thepict@height)
\put(0,0){\psfig{figure=#1,width=#2,height=#3,clip=}}
\SetScale{0.283466457}
\SetWidth{1.763889}
{#4}
\end{picture}}
}
\newcounter{pict@widthfst}
\newcounter{pict@widthscd}
\newcounter{pict@widthtot}
\newcommand{\psfigaddtwo}[7]{%
\setcounter{pict@widthfst}{1*\ratio{#2+\pict@scale/2}{\pict@scale}}
\setcounter{pict@widthscd}{1*\ratio{#2+#4+\pict@scale/2}{\pict@scale}}
\setcounter{pict@widthtot}{1*\ratio{#2+#4+#6+\pict@scale/2}{\pict@scale}}
\setcounter{pict@height}{1*\ratio{#3+\pict@scale/2}{\pict@scale}}
\setlength{\unitlength}{\pict@scale}
\hbox{\hspace{-\fill}\begin{picture}(\thepict@widthtot,\thepict@height)
\put(0,0){\psfig{figure=#1,width=#2,height=#3,clip=}}
\put(\thepict@widthscd,0){\psfig{figure=#5,width=#6,height=#3,clip=}}
\SetScale{0.283466457}
\SetWidth{1.763889}
{#7}
\end{picture}}
}
\newcommand{\psfigror}[4]{%
\setcounter{pict@width}{1*\ratio{#2+\pict@scale/2}{\pict@scale}}
\setcounter{pict@height}{1*\ratio{#3+\pict@scale/2}{\pict@scale}}
\setlength{\unitlength}{\pict@scale}
\hbox{\begin{picture}(\thepict@width,\thepict@height)
\put(0,\thepict@height){\psfig{figure=#1,width=#3,height=#2,clip=,angle=270}}
\SetScale{0.283466457}
\SetWidth{1.763889}
{#4}
\end{picture}}
}
\newcommand{\psfigrol}[4]{%
\setcounter{pict@width}{1*\ratio{#2+\pict@scale/2}{\pict@scale}}
\setcounter{pict@height}{1*\ratio{#3+\pict@scale/2}{\pict@scale}}
\setlength{\unitlength}{\pict@scale}
\hbox{\begin{picture}(\thepict@width,\thepict@height)
\put(0,0){\psfig{figure=#1,width=#3,height=#2,clip=,angle=90}}
\SetScale{0.283466457}
\SetWidth{1.763889}
{#4}
\end{picture}}
}
\catcode`\@=12 
\newlength\listtextwidth
\newcommand{\shiftfloat}[1]{
  \setlength{\listtextwidth}{\textwidth}
  \addtolength{\listtextwidth}{-#1}}
\newcommand{\none}{\hbox{---}}
\newcommand{\pcite}[1]{{\protect\cite{#1}}}
\newcommand{\pnl}{\protect{\newline}}
\catcode`\@=11 
\newlength{\@tabfninsert}
\newlength{\@tabfnwidth}
\newcommand{\tabfootnote}[2]{%
  \setlength{\@tabfninsert}{0.8em}
  \setlength{\@tabfnwidth}{\textwidth}
  \addtolength{\@tabfnwidth}{-\@tabfninsert}
  \addtolength{\@tabfnwidth}{-0.4em}
  \noindent\makebox[\@tabfninsert][r]{\footnotesize$^{#1}$\hfil}\hfill%
  \parbox[t]{\@tabfnwidth}{\footnotesize #2\hfill}}
\catcode`\@=12 

%% file: LaTeX/user/def.tex
\newcounter{FiguresAvailable}
\newlength{\halftextwidth}
\newlength{\pictwidth}
\newcommand{\PlotDir}{epsfiles}
\newcommand{\CondFigEW}[5]{%
\ifthenelse{\value{FiguresAvailable} = 1}{
\begin{figure}[h!]
\begin{center}
 \mbox{\epsfig{figure=#4,width=#5\textwidth}}
\caption[#2]{#3}
\label{fig-#1}
\end{center}
\end{figure}}%
{
\begin{figure}[h]
\caption[#2]{}
\label{fig-#1}
\end{figure}
\vspace*{-1.5ex}
\begin{center}
 \mbox{\textbf{eps file #4 to be inserted here}}
\end{center}
\vspace*{.5ex}
}}
\newcommand{\CondFigE}[4]{%
\ifthenelse{\value{FiguresAvailable} = 1}{
\begin{figure}[h!]
\begin{center}
 \mbox{\epsfig{figure=#4,width=\textwidth}}
\caption[#2]{#3}
\label{fig-#1}
\end{center}
\end{figure}}%
{
\begin{figure}[h]
\caption[#2]{}
\label{fig-#1}
\end{figure}
\vspace*{-1.5ex}
\begin{center}
 \mbox{\textbf{eps file #4 to be inserted here}}
\end{center}
\vspace*{.5ex}
}}
\newcommand{\CondFigZ}[5]{%
\ifthenelse{\value{FiguresAvailable} = 1}{
\begin{figure}[h!]
\begin{center}
\begin{tabular}{llll}
 \hspace*{2.em}\hspace*{-\halftextwidth}\mbox{\epsfig{figure=#4,
      width=0.49\textwidth}} & 
 \hspace*{-\halftextwidth} a)\hspace*{-1.0em}\hspace*{\halftextwidth} &
 \hspace*{-1.0em}\mbox{\epsfig{figure=#5,
      width=0.49\textwidth}} &
 \hspace*{-\halftextwidth} b) \\
 \end{tabular}  
\caption[#2]{#3}
\label{fig-#1}
\end{center}
\end{figure}}%
{
\begin{figure}[h]
\caption[#2]{}
\label{fig-#1}
\end{figure}
\vspace*{-2.ex}
\begin{center}
\textbf{eps files\\ 
\begin{tabular}{cc}
 #4 & #5
\end{tabular}\\
to be inserted here}
\end{center}
\vspace*{.5ex}
}}
\newcommand{\CondFigD}[6]{%
\ifthenelse{\value{FiguresAvailable} = 1}{
\begin{figure}[h!]
\begin{center}
\begin{tabular}{lclc}
 \hspace*{2.em}\hspace*{-\halftextwidth}\mbox{\epsfig{figure=#4,
      width=0.49\textwidth}} &
 \hspace*{-\halftextwidth} a)\hspace*{-1.0em}\hspace*{\halftextwidth} &
 \hspace*{-1.0em}\mbox{\epsfig{figure=#5,
      width=0.49\textwidth}} &
 \hspace*{-\halftextwidth} b) \\[.5ex]
 \hspace*{2.em}\hspace*{-\halftextwidth}\mbox{\epsfig{figure=#6,
      width=0.49\textwidth}} &
 \hspace*{-\halftextwidth} c)\hspace*{-1.0em}\hspace*{\halftextwidth} & \\
\end{tabular}  
\caption[#2]{#3}
\label{fig-#1}
\end{center}
\end{figure}}%
{
\begin{figure}[h]
\caption[#2]{}
\label{fig-#1}
\end{figure}
\vspace*{-2.ex}
\begin{center}
 \textbf{eps files\\ 
\begin{tabular}{cc}
 #4 & #5\\
 #6 &
\end{tabular}\\
to be inserted here}
\end{center}
\vspace*{.5ex}
}}
\newcommand{\CondFigRowD}[9]{%
\ifthenelse{\value{FiguresAvailable} = 1}{
\begin{figure}[h!]
\begin{center}
\begin{tabular}{ccc}
 \mbox{\epsfig{figure=#4,width=#7\textwidth}} &
 \mbox{\epsfig{figure=#5,width=#8\textwidth}} &
 \mbox{\epsfig{figure=#6,width=#9\textwidth}}
\end{tabular}\\
\begin{tabular}{lll}
 \hspace*{#7\halftextwidth} $\mathrm{(a)}$ \hspace*{#7\halftextwidth}&
 \hspace*{#8\halftextwidth} $\mathrm{(b)}$ \hspace*{#8\halftextwidth}&
 \hspace*{#9\halftextwidth} $\mathrm{(c)}$ \hspace*{#9\halftextwidth} \\
\end{tabular}
\caption[#2]{#3}
\label{fig-#1}
\end{center}
\end{figure}}%
{
\begin{figure}[h]
\caption[#2]{}
\label{fig-#1}
\end{figure}
\vspace*{-2.ex}
\begin{center}
 \textbf{eps files\\ 
\begin{tabular}{cc}
 #4 & #5\\
 #6 &
\end{tabular}\\
to be inserted here}
\end{center}
\vspace*{.5ex}
}}
\newcommand{\CondFigV}[7]{%
\ifthenelse{\value{FiguresAvailable} = 1}{
\begin{figure}[h!]
\begin{center}
\begin{tabular}{lclc}
 \hspace*{2.em}\hspace*{-\halftextwidth}\mbox{\epsfig{figure=#4,
      width=0.49\textwidth}} &
 \hspace*{-\halftextwidth} a)\hspace*{-1.0em}\hspace*{\halftextwidth} &
 \hspace*{-1.0em}\mbox{\epsfig{figure=#5,
      width=0.49\textwidth}} &
 \hspace*{-\halftextwidth} b) \\[.5ex]
 \hspace*{2.em}\hspace*{-\halftextwidth}\mbox{\epsfig{figure=#6,
      width=0.49\textwidth}} &
 \hspace*{-\halftextwidth} c)\hspace*{-1.0em}\hspace*{\halftextwidth} &
 \hspace*{-1.0em}\mbox{\epsfig{figure=#7,
      width=0.49\textwidth}} &
 \hspace*{-\halftextwidth} d) \\
\end{tabular}  
\caption[#2]{#3}
\label{fig-#1}
\end{center}
\end{figure}}%
{
\begin{figure}[h]
\caption[#2]{}
\label{fig-#1}
\end{figure}
\vspace*{-2.ex}
\begin{center}
 \textbf{eps files\\ 
\begin{tabular}{cc}
 #4 & #5\\
 #6 & #7
\end{tabular}\\
to be inserted here}
\end{center}
\vspace*{.5ex}
}}
\newcommand{\ptmiss}{\ensuremath{p_T^{\mathrm{miss}}}}
\newcommand{\scalEt}{\ensuremath{E_{\mathrm{t,s}}}}
\newcommand{\Qprisqd}{\ensuremath{{Q'}^{\hspace*{0.1em}2}}}
\newcommand{\QprisqdDA}{\ensuremath{{Q'}_\DA^{\hspace*{0.2em}2}}}
\newcommand{\Qprisqdtr}{\ensuremath{{Q'}_{\mathrm{gen}}^{\hspace*{0.2em}2}}}
\newcommand{\Qprisqdmin}{\ensuremath{{Q'}_{\mathrm{min}}^{\hspace*{0.2em}2}}}
\newcommand{\regmeth}{region method }
\newcommand{\RegMeth}{Region Method }
\newcommand{\gen}{{\rm gen}}
\newcommand{\Hwg}{{\rm Hwg}}
\newcommand{\Dja}{{\rm Dja}}
\newcommand{\EtaInst}{\ensuremath{\eta_{\mathrm{Inst}}}}
\newcommand{\MInst}{\ensuremath{M_{\mathrm{Inst}}}}
\newcommand{\PtJet}{\ensuremath{p_T^{\mathrm{Jet}}}}
\newcommand{\lPtJet}{\ensuremath{\mathrm{log}_{\mathrm{10}}\ p_T^{\mathrm{Jet}}}}
\newcommand{\lPtInst}{\ensuremath{\mathrm{log}_{\mathrm{10}}\ P_t(\mathrm{Inst})}}
\newcommand{\GammaHad}{\ensuremath{\gamma_{\mathrm{h}}}}
\newcommand{\EpsP}{\ensuremath{{\epsilon'}}}
\newcommand{\Circ}{\ensuremath{C}}
\newcommand{\Sph}{\ensuremath{S}}
\newcommand{\NzufosI}{\ensuremath{N_{\mathrm{EFO}}}}
\newcommand{\Nzuftr}{\ensuremath{N_{\mathrm{EFT}}}}
\newcommand{\Ets}{\ensuremath{E_{\mathrm{t,s}}}}

%% file: InstPaper-cit.tex
\def\citeBelavin{{\cite{%
pl:b59:85%
}}\xspace}
\def\citeNameInst{{\cite{%
prl:37:8,*pr:d14:3432,*prep:142:357%
}}\xspace}

\def\citeElWeakInst{{\cite{%
pl:b188:506,*np:b330:1,*np:b343:310,*np:b365:3,*zfp:c66:285,*pl:b555:227,proc:quarks:1994:170%
}}\xspace}
\def\citeShortDist{{\cite{%
pl:b314:237,proc:quarks:1994:170,np:b507:134,pl:b459:249%
}}\xspace}

\def\citeQCDINS{{\cite{%
cpc:132:267,*proc:dis:1995:341%
}}\xspace}
\def\citeHOneInst{{\cite{%
epj:c25:495%
}}\xspace}
\def\citeFisherAlg{{\cite{%
ae:7:179,bllo:1998:datenanalyse%
}}\xspace}
\def\citeInstPhen{{\cite{%
proc:quarks:1994:170,np:b507:134,pl:b438:217,pl:b459:249,cpc:132:267,prl:b503:331%
}}\xspace}
\def\citeInstChar{{\cite{%
proc:quarks:1994:170%
}}\xspace}
\def\citeFidReg{{\cite{%
pl:b459:249%
}}\xspace}
\def\citeNonPlanar{{\cite{%
np:b507:134%
}}\xspace}

\def\citeZEUSDet{{\cite{%
zeus:1993:bluebook%
}}\xspace}
\def\citeCTD{{\cite{%
nim:a279:290,*npps:b32:181,*nim:a338:254%
}}\xspace}
\def\citeCAL{{\cite{%
nim:a309:77,*nim:a309:101,*nim:a321:356,*nim:a336:23%
}}\xspace}
\def\citeLumiDet{{\cite{%
desy-92-066%
}}\xspace}
\def\citeTrigger{{\cite{%
zeus:1993:bluebook,proc:com:92:222%
}}\xspace}
\def\citeDjango{{\cite{%
cpc:81:381%
}}\xspace}
\def\citeCDM{{\cite{%
np:b306:746,*pl:b175:453,*zfp:c43:625%
}}\xspace}
\def\citeAriadne{{\cite{%
cpc:71:15%
}}\xspace}

\def\citeStringFrag{{\cite{%
prep:97:31%
}}\xspace}
\def\citeJetset{{\cite{%
cpc:82:74,*cern-th:7112:93%
}}\xspace}
\def\citeHeracles{{\cite{%
cpc:69:155%
}}\xspace}
\def\citeCTEQfour{{\cite{%
pr:d55:1280%
}}\xspace}
\def\citeHerwig{{\cite{%
np:b310:461,*cpc:67:465%
}}\xspace}
\def\citeLepto{{\cite{%
cpc:101:108%
}}\xspace}
\def\citeClusterHad{{\cite{%
np:b238:492%
}}\xspace}

\def\citeGeant{{\cite{%
cern-dd:84:1%
}}\xspace}

\def\citeZufos{{\cite{%
brisk:phd:1998,*epj:c1:81%
}}\xspace}
\def\citeSinistra{{\cite{%
nim:a365:508,*nim:a391:360%
}}\xspace}
\def\citeDAelecRec{{\cite{%
proc:hera:1991:23,*proc:hera:1991:43%
}}\xspace}
\def\citeJBRec{{\cite{%
proc:epfacility:1979:391%
}}\xspace}
\def\citektclus{{\cite{%
np:b406:187%
}}\xspace}
\def\citeInclktclus{{\cite{%
pr:d48:3160%
}}\xspace}

\def\citeSteepDecreaseAtLowXpri{{\cite{%
proc:dis:2000:318,pl:b459:249,prl:b503:331%
}}\xspace}

\def\citeCarlKo{{\cite{%
hep-ph:00%
}}\xspace}
\def\citeEtamax{{\cite{%
pl:b421:368%
}}\xspace}
\def\citeshillert{{\cite{%
thesis:hillert:2002%
}}\xspace}

%% file: InstPaper-tit.tex
\title{
\begin{flushright}
\normalsize {DESY 03-201 \\
December 2003} 
\vspace{1cm}
\end{flushright}
Search for QCD-instanton induced events 
 in  deep inelastic ep 
scattering at HERA
}                                                       
                    
\author{ZEUS Collaboration}
\abstract{

A search for QCD-instanton-induced events in deep inelastic $ep$ scattering
has been performed with the ZEUS detector at the HERA collider, using data
corresponding to an integrated luminosity of $38\pbi$. 
A kinematic range defined by cuts on the photon virtuality, $Q^{2} > 120
 \gev^2$,
and on the Bjorken scaling variable, $x > 10^{-3}$, has been investigated.
The QCD-instanton induced events were modelled by the Monte Carlo generator
 QCDINS.
A background-independent, conservative 95\% confidence level  upper limit for
 the instanton cross section of $26 \pb$
is obtained, to be compared with the theoretically expected value of
 $8.9 \pb$.
}

\makezeustitle

%% file: auth113_out.tex
\def\3{\ss}                                                                                        
\pagenumbering{Roman}                                                                              
                                                   %
\begin{center}                                                                                     
{                      \Large  The ZEUS Collaboration              }                               
\end{center}                                                                                       
  S.~Chekanov,                                                                                     
  M.~Derrick,                                                                                      
  D.~Krakauer,                                                                                     
  J.H.~Loizides$^{   1}$,                                                                          
  S.~Magill,                                                                                       
  S.~Miglioranzi$^{   1}$,                                                                         
  B.~Musgrave,                                                                                     
  J.~Repond,                                                                                       
  R.~Yoshida\\                                                                                     
 {\it Argonne National Laboratory, Argonne, Illinois 60439-4815}, USA~$^{n}$                       
\par \filbreak                                                                                     
  M.C.K.~Mattingly \\                                                                              
 {\it Andrews University, Berrien Springs, Michigan 49104-0380}, USA
% ...
\newpage                                                                                           
$^{\    1}$ also affiliated with University College London, London, UK \\                          
% ...
                                                           %
                                                           %
\newpage   
                                                           %
                                                           %
\begin{tabular}[h]{rp{14cm}}                                                                       
$^{a}$ &  supported by the Natural Sciences and Engineering Research                               
          Council of Canada (NSERC) \\                                                             
$^{b}$ &  supported by the German Federal Ministry for Education and                               
          Research (BMBF), under contract numbers HZ1GUA 2, HZ1GUB 0, HZ1PDA 5, HZ1VFA 5\\         
$^{c}$ &  supported by the MINERVA Gesellschaft f\"ur Forschung GmbH, the                          
          Israel Science Foundation, the U.S.-Israel Binational Science                            
          Foundation and the Benozyio Center                                                       
          for High Energy Physics\\                                                                
$^{d}$ &  supported by the German-Israeli Foundation and the Israel Science                        
          Foundation\\                                                                             
$^{e}$ &  supported by the Italian National Institute for Nuclear Physics (INFN) \\                
$^{f}$ &  supported by the Japanese Ministry of Education, Culture,                                
          Sports, Science and Technology (MEXT) and its grants for                                 
          Scientific Research\\                                                                    
$^{g}$ &  supported by the Korean Ministry of Education and Korea Science                          
          and Engineering Foundation\\                                                             
$^{h}$ &  supported by the Netherlands Foundation for Research on Matter (FOM)\\                   
$^{i}$ &  supported by the Polish State Committee for Scientific Research,                         
          grant no. 620/E-77/SPB/DESY/P-03/DZ 117/2003-2005\\                                      
$^{j}$ &  partially supported by the German Federal Ministry for Education                         
          and Research (BMBF)\\                                                                    
$^{k}$ &  partly supported by the Russian Ministry of Industry, Science                            
          and Technology through its grant for Scientific Research on High                         
          Energy Physics\\                                                                         
$^{l}$ &  supported by the Spanish Ministry of Education and Science                               
          through funds provided by CICYT\\                                                        
$^{m}$ &  supported by the Particle Physics and Astronomy Research Council, UK\\                   
$^{n}$ &  supported by the US Department of Energy\\                                               
$^{o}$ &  supported by the US National Science Foundation\\                                        
$^{p}$ &  supported by the Polish State Committee for Scientific Research,                         
          grant no. 112/E-356/SPUB/DESY/P-03/DZ 116/2003-2005,2 P03B 13922\\                       
$^{q}$ &  supported by the Polish State Committee for Scientific Research,                         
          grant no. 115/E-343/SPUB-M/DESY/P-03/DZ 121/2001-2002, 2 P03B 07022\\                    
\end{tabular}                                                                                      
                                                           %
                                                           %

%% file: InstPaper-txt.tex
\input{zeus_defs}
\input{defs}
\pagenumbering{arabic}
\pagestyle{plain}
\section{Introduction}
\label{sect-Intro}
In the Standard Model, both the strong and the electroweak interactions
are described by non-Abelian gauge theories.
In such theories, the ground state has a rich topological structure,
associated with non-perturbative fluctuations of the
gauge fields \citeBelavin, called instantons \citeNameInst.
They can be interpreted as tunneling processes of the gauge fields
between topologically distinct types of vacuum states. 

Although the existence of instantons is required by the Standard Model, 
they have not been observed.
While in electroweak interactions instantons are predicted to play a role
 only at 
centre-of-mass energies $\gg 10 \tev$ \citeElWeakInst,
 QCD instanton effects are expected to become sizeable at much lower 
energies, where they
 are predicted to have short-distance implications
\citeShortDist.
In particular, they can induce characteristic events in  
 deep inelastic scattering (DIS). The HERA $ep$ collider
 offers a unique opportunity to discover instantons. Their discovery  
would constitute a confirmation of an essential non-perturbative Standard
 Model prediction, connected with the QCD vacuum.
Results of such a search have recently been reported by the H1
collaboration \citeHOneInst. 

Since instanton-induced events are predicted to contribute less than 1\%
 to the cross section of the neutral current DIS
sample used in this search, it is essential to find variables that
efficiently discriminate between instanton and standard DIS events.
  Statistical discrimination methods have been employed to obtain
event samples with a larger fraction of instanton events in a search for a
 possible signal.


\section{Experimental setup}
\label{sect-det}
%
The data sample used in this search corresponds to an integrated luminosity of
$38.3\pm0.6\pbi$ collected with the ZEUS detector at the HERA collider.
During the years 1996 and 1997, positrons
of energy $E_e = 27.5 \gev$ collided with protons of energy $E_p =
 820\gev$.
A detailed description of the ZEUS detector can be found elsewhere 
\citeZEUSDet.
The main detector components used in the search presented here are the central 
tracking detector (CTD) \citeCTD, operating in a magnetic field of 
$1.43\, \mathrm{T}$ provided by a thin superconducting solenoid, and the 
uranium-scintillator calorimeter (CAL) \citeCAL.

Tracking information is provided by the CTD, in which the momenta of tracks in
 the polar-angle%
\footnote{\ZcoosysB}
region $15^{\circ} < \theta < 164^{\circ}$ are reconstructed. The CTD consists
of 72 cylindrical drift chamber layers, organised in nine superlayers. The
relative transverse-momentum resolution for full-length tracks is
 $\sigma(p_T)/p_T = 0.0058 p_T \oplus 0.0065 \oplus 0.0014/p_T$,
 with $p_T$ in $\mathrm{GeV}$.

The CAL covers 99.7\% of the total solid angle. It is divided into
 three parts with a corresponding division in $\theta$, as viewed from the
nominal interaction point:
forward (FCAL, $2.6^{\circ} < \theta < 36.7^{\circ}$), barrel (BCAL,
 $36.7^{\circ} <
\theta < 129.1^{\circ}$),  and rear (RCAL, $129.1^{\circ} < \theta <
 176.2^{\circ}$).
Each of the CAL parts is subdivided into towers, which in turn are
segmented longitudinally
 into one electromagnetic (EMC) and one (RCAL)  or two (FCAL, BCAL) hadronic
(HAC) sections.
The smallest subdivision of the CAL is called a cell.
Under test-beam conditions, the CAL single-particle relative energy resolution
was $ \sigma{(E)}/E= 0.18 / \sqrt {E}$ for electrons and $ \sigma{(E)}/E=
0.35 / \sqrt{E}$ for hadrons, with $E$ in $\mathrm{GeV}$.

The luminosity was measured using the Bethe-Heitler reaction $ep \rightarrow
e \gamma p$. The resulting small-angle energetic photons were measured by
the  luminosity monitor \citeLumiDet, a lead-scintillator calorimeter
placed  in the HERA tunnel at $Z = - 107\,\mathrm{m}$. 
\section{Characteristics of instanton-induced events}
\label{sect-InstIndEvts}
Ringwald
and Schrempp \citeInstPhen have identified kinematic regions in DIS
that allow a perturbative calculation of instanton-induced processes.
These processes lead to a characteristic final state, which may allow  
instanton-induced events to be distinguished from
 standard DIS processes. %
Figure~\ref{fig-InstIndEvtKine} shows a diagram of an instanton-induced
event in an $ep$ collision. %
The incoming lepton emits a photon, with four-vector $q$, which in turn
 transforms into a quark-antiquark pair. %
One of these quarks hadronises to form the current jet with four-vector $q''$. %
The other quark, with four-vector $q'$, fuses with a gluon (four-vector $g$) 
from the proton in the presence of an instanton. %
The phenomenological characteristics of instanton-induced events
 can be summarised as follows \citeInstChar :
in the hard subprocess 
exactly one $q \bar{q}$ pair of each of the $n_f$
 kinematically accessible quark flavours participates in the
quark gluon fusion process, either as incoming or outgoing fermion line.
This gives rise to a high-multiplicity
final state. The produced particles are expected to be isotropically
distributed in their centre-of-mass frame. In addition, the events have large
transverse energy in the hadronic centre-of-mass frame.

Instanton-induced events in DIS were simulated using 
the Monte Carlo (MC) generator QCDINS 2.0 \citeQCDINS, which 
 simulates the hard subprocess in the presence of an 
instanton.
For the description of parton showers and hadronisation,
 HERWIG 5.9 \citeHerwig is used.
%
The simulation of the hard subprocess is accomplished by applying
instanton perturbation theory around the one-instanton solution.
By comparing the results obtained from instanton perturbation theory
to those from lattice simulations of QCD, the fiducial
region of the QCDINS MC program has been derived \citeFidReg
in terms of the kinematic variables 
$\Qprisqd = - {q'}^2 = - (q - q'')^2$, $ x' = {Q'}^2/(2 g \cdot q')$ and
$Q^2=-q^2$, 
 describing the 
hard subprocess of instanton-induced events:
%
\[ \begin{array}{c} 
       Q' \geq {Q'}_{\mathrm{min}} = 30.8 \cdot 
         \Lambda^{(n_{f})}_{\overline{\mathrm{MS}}} =
       \sqrt{113\gev^2}\hspace*{3em};\quad 
        x' \geq {x'}_{\mathrm{min}} = 0.35, \\[2.5ex]
     \end{array}
\]
where $\Lambda^{(n_{f})}_{\overline{\mathrm{MS}}}$ is the QCD scale. 

For QCDINS, cuts on the generated $x'$ and ${Q'}^2$ variables were made at the
values given above in order to restrict the sample to the region where the 
calculation is reliable.
Non-planar contributions \citeNonPlanar, which are not taken into account in
 the 
calculation, were suppressed by a cut on the photon virtuality, 
$  Q^2 \geq Q^2_{\mathrm{min}} = {Q'}^2_{\mathrm{min}}$.

The generated events were passed through the GEANT 3.13 -based \citeGeant ZEUS
detector- and trigger-simulation programs.
 They were
reconstructed and analysed using the same program chain as the data.
%
\section{Simulation of standard DIS events}
\label{sect-nDISEvts}
Standard DIS events were generated using the LEPTO 6.5 program \citeLepto
interfaced to HERACLES \citeHeracles via
 DJANGOH 1.1 \citeDjango. The HERACLES program includes first-order 
electroweak radiative corrections.
The CTEQ4 \citeCTEQfour proton parton distribution functions (PDF) were used.
The QCD radiation was modelled with the
colour-dipole model (CDM) \citeCDM by using the ARIADNE 4.10 program
 \citeAriadne.
Fragmentation into hadrons was performed using the Lund string model
\citeStringFrag as implemented in JETSET \citeJetset.
 In order to improve the description of the sphericity distribution in the
 hadronic final state (see Section~\ref{sect-InstRec}),
the width of the transverse momentum distribution of primary hadrons
 (i.e. JETSET parameter PARJ(21)) was lowered to 
$0.28 \gev$.
The diffractive contribution to the 
neutral current sample was taken into account by adding $12\%$ of diffractive
 RAPGAP events.
This percentage was determined from a fit of the distribution of the
 variable $\eta_{\mathrm{max}}$ \citeEtamax,
which is the pseudorapidity of the calorimeter
energy deposit with the smallest polar angle and an energy above
$400 \mev$. 
In what follows, the term DJANGOH always refers to the combination of
DJANGOH and RAPGAP.
For a cross-check of the results, the generator HERWIG 5.9 \citeHerwig has 
been used; this program also provides
the description of parton showers and hadronisation for the QCDINS
MC. In HERWIG, coherence effects in the final-state cascade are
included by angular ordering of successive parton emissions, and a clustering
model is used for hadronisation \citeClusterHad.
Electroweak radiative effects are not included in HERWIG.
Detector resolution and selection efficiency were simulated as for the
signal sample (see Section~\ref{sect-InstIndEvts}).

%
%
\section{Event Selection and Reconstruction}
\label{sect-EvtSel}
\subsection{Reconstruction of kinematic variables}
\label{subs-EvtRec}
Both track and calorimeter information were used for event reconstruction.
Calorimeter cells were first grouped to form clusters which were
then associated with tracks, where possible, to form energy-flow objects
(EFOs)~\citeZufos.
The hadronic final state of an event comprises all EFOs that do not stem
from the scattered positron.

Positron identification was based on the pattern of energy deposits in
the CAL \citeSinistra. The positron energy $E_{\mathrm{DA}}$ was calculated
using the double angle (DA) method \citeDAelecRec. Tracking information
was used to determine the positron polar angle,  
if the polar angle in the CAL was above $0.3$ rad and if 
the track traversed more than three CTD superlayers.
Otherwise these angles were determined from the CAL information.

The kinematic region investigated was defined by cuts on
 $Q^2$, and on the Bjorken scaling variables
$x$ and $y$. 
%
The variables $Q^2$ and $x$ were reconstructed using the 
DA method, and 
 $y$ was reconstructed using the Jacquet-Blondel method \citeJBRec.

In order to reconstruct the kinematic variable $\Qprisqd$ (see
 Fig.~\ref{fig-InstIndEvtKine}),
EFOs were assigned to the current jet or to the instanton part of the 
hadronic final state.
The current jet was identified by applying the $k_T$-cluster algorithm \citektclus 
in the longitudinally invariant mode \citeInclktclus 
on all EFOs in the hadronic centre-of-mass frame (hcms). 
The photon direction was chosen as the negative 
$Z$ direction.
Monte Carlo studies showed, that
in a region where the cross section for instanton related events is
 enhanced, and for high transverse jet momenta,
the current quark has, on average, a 
smaller pseudorapidity than the partons assigned to the instanton. 
 The current jet was therefore found 
 as follows: 
a list of jets with a pseudorapidity 
$\eta_{\mathrm{jet}}^{\mathrm{hcms}}$ less than the
 transverse-momentum-weighted mean pseudorapidity of all EFOs in the hadronic final state
 was made ($\eta_{\mathrm{jet}}^{\mathrm{hcms}} <
\sum_{\mathrm{EFOs}} \eta \cdot p_T / \sum_{\mathrm{EFOs}} p_T$).
%
%
%
Here the jet pseudorapidity 
was calculated from the jet four-momentum, obtained by summing the 
four-momenta of all the EFOs assigned to the jet.
Of the jets in the list, the one with the 
highest transverse momentum was chosen as the current jet.

The variable ${Q'}^2_\DA$ was calculated using the expression
\[{Q'}^2_\DA = Q^2_\DA + \frac{W^2_\DA - Q^2_\DA}{W_\DA} \cdot
\sum_{i\in {\mathrm{jet}}} E_i -
\frac{W^2_\DA + Q^2_\DA}{W_\DA} \cdot \sum_{i\in {\mathrm{jet}}} p_{Z,i}
-M^2_{\mathrm{jet}}\ \,, \]
with
\[ W_\DA = \sqrt{Q^2_\DA (1 - x_\DA)/x_\DA + m^2_p}\ \,, \]
where $M_{\mathrm{jet}}$ is the current jet mass and $m_p$ the proton mass.
According to MC studies,
 ${Q'}^2$ is reconstructed with a relative accuracy of about 30\%. However,
 the distribution of
 the reconstructed ${Q'}^2$ has a long tail in the direction of
 overestimation of
 the true value.
%
%
\subsection{Selection of neutral current DIS events}
\label{subs-NCDISsel}
A three-level trigger \citeTrigger was used to select events
 online.
A high-$Q^2$ neutral current DIS sample was selected requiring, at the third
level trigger, a positron with an 
energy greater than $4 \gev$ at a position outside a radius of $25 \cm$ 
on the calorimeter.

In order to select neutral current DIS events offline, the following cuts were
 applied, defining  a fiducial sample:%
%
%
\begin{tabbing}
\hspace*{0em}\=\ \hspace*{1em}\=\ \hspace*{0.5em}\=\kill
%
\>{$\bullet$ kinematic cuts: $Q^2_{\DA} > 120 \gev^2$, $x_{\DA} > 10^{-3}$,
$y_{\JB} > 0.05$};\\[1.ex]
\>{$\bullet$ cuts to ensure the quality of the %
positron reconstruction:}\\[1.ex]
\>\>{$\circ$ The radial position of the positron track impact point on the rear
     calorimeter}\\
\>\>\>{ surface was required to exceed 36 cm;}\\
\>\>{$\circ$ $E_{\mathrm{DA}} > 10 \gev$;}\\[.5ex]
\>\>{$\circ$ $E_{\mathrm{cne}} < 5 \gev$, where $E_{\mathrm{cne}}$ is the
         energy, not associated with the positron, }\\
\>\>\>{  found inside a cone having an $(\eta,\phi)$-radius $R =
       \sqrt{(\Delta \phi)^2 + (\Delta\eta)^2} = 0.8$ }\\
\>\>\>{around the track of the positron candidate;}\\
[2.5ex]
\>{$\bullet$ suppression of photoproduction events:}\\[1.ex]
\>\>{$\circ$ $y_{\mathrm{el}} < 0.90$, where $y_{\mathrm{el}}$ is calculated
      from the scattered electron;  }\\[.5ex]
\>\>{$\circ$ $35 \gev < \sum_{\mathrm{EFOs}} (E - p_{Z}) < 65 \gev$, where the
         sum runs over the energy- }\\
\>\>\>{and $Z$-components of all EFO four-momenta;}\\[0.5ex]
\>\>{$\circ$ $\mathrm{DCA} < 10 \cm$, where DCA is the distance of closest 
        approach of the}\\
\>\>\>{ positron track to the centre of the cluster of CAL
        cells assigned to it;}\\[2.5ex]
%
\>{$\bullet$ vertex cut: $Z$ position of the event vertex, 
        $\left| Z_{\mathrm{VTX}} \right| < 50\cm$, consistent with an
        $ep$}\\
\>\>{ interaction;  }\\[1.ex]
%
%
\>{$\bullet$ restriction of the data sample to a region, where the QCDINS
      MC calculation is}\\
\>\>{reliable: $Q'^2_{\DA} > 140 \gev^2$.}\\[.5ex]
\end{tabbing}
Figure~\ref{fig-QPR} shows the measured $\QprisqdDA$ distribution, compared to
various MC predictions. There is
 agreement
between data and the standard DJANGOH DIS MC sample at a level of about
 10 \%. A
larger discrepancy is seen between the data and the HERWIG MC sample. Also
 shown
is the prediction of the QCDINS MC program, normalised to the number of events
in the data. It has a very different distribution from both the data and the
standard DIS events.

No cut was made on the variable $x'$ for the standard DIS MC sample and the
 data because $x'$ cannot be well reconstructed.
On the other hand, the QCDINS Monte Carlo sample includes a cut on the
 generated variable  $x'>0.35$.
Lattice calculations show a steep decrease of the instanton
contribution towards small values of $x'<0.35$, corresponding to a small 
separation between instantons and anti-instantons, 
 suggesting that this region can be neglected \citeSteepDecreaseAtLowXpri.

Application of the above cuts resulted in a sample of $91846$ events for the
 data. Normalised to the data luminosity,
 QCDINS predicts $578$ events, DJANGOH predicts $88300$ and HERWIG
predicts $76400$ events. 
The corresponding predicted instanton cross section is $8.9\pb$. 
The statistical uncertainties on these numbers are negligible compared to
 the uncertainty on the luminosity.

The discrepancy in the number of events predicted by
HERWIG and DJANGOH
can be traced to the cut in $Q'^2$, and arises from differences in the $Q'^2$
reconstruction due to the different hadronisation models in the MC
programs.
 Without this cut, the numbers agree
within the estimated uncertainty. 

In the subsequent analysis, numbers of standard DIS Monte Carlo events 
were normalised to the number of data events in the fiducial 
sample, where the predicted QCDINS contribution is negligible 
($\approx 0.7\%$).
\section{Definition of discriminating variables}
\label{sect-InstRec}
Two kinds of discriminating variables have been considered:
those connected with the kinematics of an 
instanton-induced event and those connected with the
final-state particles of the instanton system, the so-called shape
variables. The kinematic variables chosen were  $\QprisqdDA$, 
(see Fig.~\ref{fig-InstIndEvtKine} and Section~\ref{sect-InstIndEvts})
and $p_T^{\mathrm{jet}}$, the transverse momentum of the current jet in the   
 hcms.

The shape variables were calculated from  a subset of 
hadronic final state EFOs  assigned to the
instanton, referred to as the ``instanton region'' in what
follows. 
The instanton region comprises all EFOs which were not assigned to the
current jet and lay in the hemisphere opposite to the outgoing
proton remnant, i.e. polar angle $\theta^{\mathrm{hcms}}_{\mathrm{EFO}} >
 90^{\circ}$.
Once the instanton region was
 identified, the following shape variables were calculated:
\begin{itemize}
\item $\NzufosI$, the multiplicity of EFOs;

\item  $\Nzuftr$, the multiplicity  of tracks, that were 
    used in constructing these EFOs;

\item  $C$, the circularity;
this is a measure of the isotropy
of EFOs
in the hcms with respect to
the photon-proton axis.
To determine the circularity, the normalised two-dimensional 
momentum tensor, 
\[ M^{(\mathrm{2D})}_{\alpha\beta} = 
        \frac{\sum_j p_{j,\alpha} p_{j, \beta}}%
        {\sum_j \left(p^2_{j,X} + p^2_{j,Y}\right)} \ \ \ %
        \mathrm{with}\ \alpha, \beta = X,\, Y \  \]

was computed in the hcms from the three-momenta of the EFOs, requiring
the number of EFOs to be larger than two.  
From its eigenvalues $\lambda_1$, $\lambda_2$, with 
$\lambda_1 > \lambda_2$, the circularity $C$ was then obtained:
 $C = 2 (1-\lambda_1)$.\\

\item $S$, the sphericity;
this is a measure of how isotropically
a collection of three-momenta is distributed. Large
values correspond to a more isotropic distribution.
The normalised momentum tensor, calculated from the
EFOs in their centre-of-mass frame, 
\[ M^{(\mathrm{3D})}_{\alpha\beta} = 
       \frac{\sum_j p_{j,\alpha} p_{j, \beta}}
        {\sum_j \left(p^2_{j,X} + p^2_{j,Y} + p^2_{j,Z}\right)} \ \ \ %
        \mathrm{with}\ \alpha, \beta = X,\, Y,\, Z\ \ \ , \]
has eigenvalues $Q_1, Q_2, Q_3$ with %
$0\ \leq\ Q_1\ \leq\ Q_2\ \leq\ Q_3$.
The sphericity is defined as $S$ = 3/2 $\cdot \left(Q_1 +
 Q_2\right)$. The number of EFOs for this calculation had to be larger
 than two.\\

\item  $\EpsP$, a 
measure of the density of the pseudorapidity $\eta$ of EFOs in the hcms;
to calculate $\EpsP$, the EFOs were sorted with respect to the pseudorapidity
 $\eta$ in the hcms, yielding 
$\NzufosI$ values $\eta_1$, $\eta_2$, \dots, $\eta_{\NzufosI}$.
%
The closed interval $\left[\eta_i, \eta_{i+k} \right]$, 
$i + k\, \leq\, N_{\mathrm{EFO}}$ thus contains $k + 1$ EFOs.
The variable $\EpsP$ is then defined by $\EpsP =\epsilon_c - b (\NzufosI -N_0) 
\ \ \mathrm{with} \ \ b = 0.339, \, N_0=30$ and 
\[
\epsilon_c=\frac{k}{\NzufosI -k} \sum_{i=1}^{\NzufosI - k} \frac{1}{
\eta_{i+k} - \eta_i}\ \ \ \mathrm{where}\ \ \ %
k = \bigg\{\begin{array}{ll}
        N_{\mathrm{EFO}} / 2 & \ \ \ 
        \mathrm{for\ }N_{\mathrm{EFO}}\ \mathrm{even,}\\
        (N_{\mathrm{EFO}} + 1) / 2 & \ \ \ 
        \mathrm{for\ }N_{\mathrm{EFO}}\ \mathrm{odd.}\\
        \end{array}\ \ \ 
 \]
A more detailed description and justification of the numbers $b$ and $N_0$
 can be found elsewhere \citeshillert.
\end{itemize}

Figures \ref{fig-IEvarsOne} and \ref{fig-ShapeVars} show the
 distributions of the discriminating variables
in the fiducial sample (see Section~\ref{sect-EvtSel}). The contribution of
 instanton-induced events 
is 0.7 \% according to the predictions of the QCDINS calculation. None of the
 variables have been corrected for detector or
trigger effects.

In general, there is a qualitative agreement between the shape of the data and
the DIS MC samples DJANGOH and HERWIG. The QCDINS predictions, normalised
to the number of data events, show quite different distributions, indicating
the suitability of these variables for separating instanton induced events
from background.

However, the data are not reproduced in detail by the DIS MC simulations.
 Moreover,
the two DIS MC descriptions differ from each other by a similar degree
as from the data. For the DJANGOH sample, a parameter in the JETSET
fragmentation program was tuned, as described in Section~\ref{sect-nDISEvts},
 so that the sphericity distribution of the
 data is reproduced, as shown
in Fig.~\ref{fig-ShapeVars}a. The description of the other variables,
however, as shown in the figures, is not improved.

Given the uncertainties of the DIS MC predictions, and the smallness
of the instanton contribution, it is not possible to make a reliable 
background subtraction. 
%
\section{Instanton enhancement}
\label{sect-MethIE}
Several enhancement methods were compared to each other using the following
 criteria:

From the numbers 
\[  \begin{array}{ll}
   N_O (I_O): & \mbox{number of standard DIS MC (QCDINS) events in the fiducial
 sample and}\\[1.ex]
   N_E (I_E): & \mbox{number of standard DIS MC (QCDINS) events in the
   instanton enhanced}\\[1.ex]
    & \mbox{sample,}\\[1.ex]
   \end{array}
\]
the efficiencies $r_I$, of QCDINS and $r_N$, of the standard DIS MC 
samples were obtained according to $r_I=I_E/I_O$ and $r_N=N_E/N_O$.
Values of the separation power, $P_s=r_I/r_N$, were then compared 
for different samples of similar QCDINS efficiency.  

The enhancement methods investigated \citeshillert comprised a combination of 
one-dimensional cuts on the discriminating variables, an optimised 
choice of two-dimensional cuts and the Fisher algorithm \citeFisherAlg, which 
performed best.

The Fisher algorithm was used to separate standard DIS
 and instanton events by
cutting on a linear combination of the variables $x_i$.
For $n$ input variables $x_i$, $i = 1,\,2,\,\dots, n$, the mean values 
$\overline{x_i}^s$ for the signal and $\overline{x_i}^b$ for the background 
were determined.
The correlation matrices between the variables were calculated 
according to
\[ V_{ik,s} = %
\frac{1}{N} \sum_{\mathrm{events}} 
(x_i^s - \overline{x_i}^s) (x_k^s - \overline{x_k}^s) \]
for the signal sample, and equivalently for the background ($V_{ik,b}$).
These were averaged, $\overline{V}_{ik} = \frac{1}{2} (V_{ik,s} + V_{ik,b})$,
and the resulting matrix was inverted.
The Fisher discriminant is defined as
\[ t =\sum_{i=1}^{n} w_i x_i \ \ \ , \ \ \ %
\mathrm{with} \ w_i = \sum_k (\overline{V}^{-1})_{ik} 
          (\overline{x_k}^s - \overline{x_k}^b)\ \ \ . \]

Carli and Koblitz \citeCarlKo have proposed a strategy to look for
instanton-induced processes, using variables resulting from a careful
optimisation procedure. An analysis with their variables, using the
Fisher algorithm for signal enhancement, was also carried out. In addition,
an analysis was carried out with a combination of their variables and
those of Section~\ref{sect-InstRec}. No significant improvement, as measured
by the separation power $P_s$, was achieved.

\section{Background-independent limits}
\label{subs-CutAppr}

In order to be independent of the standard DIS MC prediction, a conservative
 upper
 limit 
was set by assuming that all observed events after instanton-enhancement cuts
are signal, i.e. the standard DIS background 
was zero. 

 To derive limits on the cross section, the QCDINS
MC program was assumed to give a correct description of instanton-induced
events. 
For signal enhancement the Fisher algorithm was used, with the QCDINS sample
as signal and the DJANGOH sample as background. In this procedure, the
DJANGOH background sample is only used to determine a good discriminating
function; a non-optimal choice of background MC events will imply only that
the instanton sensitivity of the measurement is not the highest
 achievable in principle.

The Fisher discriminant was calculated from all six discriminating variables
 described in Section~\ref{sect-InstRec}
($S$, $C$, $\log(p_T^{\mathrm{jet}})$,$\NzufosI$, $\Nzuftr$, $\EpsP$).
In addition, a cut $\QprisqdDA < 250\gev^2$ was imposed, to improve instanton
enhancement.

The distribution of the Fisher discriminant $t$ for the data, the QCDINS
signal sample and the DJANGOH background sample is shown in
 Fig.~\ref{fig-FisherDiscriminant}. The distributions for the DIS MC
 simulation and
the data are rather similar, with the curve of the instanton signal being well
separated. The distribution normalised to the
predicted instanton fraction is also shown.

The numbers of data events above various values of the discriminating
 variable $t$ are shown in Table~\ref{tab-CutResults}. 
Also included are the numbers for the standard DIS
MC samples. These are slightly larger than
the number of data events. This is not surprising, since 
extremely restrictive cuts have been
chosen, which only keep events in the tails of the
standard DIS distributions. Problems are therefore expected in the modelling
of the data. 

The number of data events kept by each cut, $N_d$,
was then compared with the 
theoretically predicted number of instanton events, $N_{\rm th}$. 
The ratio $N_d / N_{\rm th}$ is equal to the ratio of
 cross-sections $R=\sigma_d / \sigma_{\rm th}$, where $\sigma_{\rm th}=8.9 \pb$
is the theoretically predicted instanton cross section for the cuts of
the fiducial sample, and $\sigma_d$ is the instanton cross section computed
 for the same
cuts, assuming that all data events are signal. This is true
under the assumption that, with zero background, the acceptances of the
 observed events and of the QCDINS MC  sample are equal. 
Figure~\ref{fig-Factor} shows the 
ratio $R=\sigma_d / \sigma_{\rm th}$,
as a function of $r_I$, the fraction of 
instanton-induced events remaining in the sample after various cuts in
the Fisher $t$ variable.

In addition to the statistical uncertainties, uncertainties on the ratio $R$ 
were taken into account by considering a $\pm 3\%$ change in the CAL energy
 scale, a change in the definition of the instanton region (see
Section~\ref{sect-InstRec}), 
and using ${Q}^2$ computed by the scattered electron variables instead of
$Q^2_{\mathrm{DA}}$.

Another uncertainty stems from the cuts made on the
variables $Q'^2$ and $x'$ by the
 QCDINS MC program at event generation,
$x' > 0.35$ and ${Q'}^2 > 113 \gev^2$.
These variables cannot be reconstructed well in the data. Therefore, the data
might include some instanton-induced events with real 
$x'$, ${Q'}^2$ values below these cut values. If, with the help of an accurate
reconstruction of $Q'^2$ and $x'$, these events could be
removed from the data sample, it would lower 
the ratio $R$, yielding a value closer to the theoretical prediction.

From the value of $R$ and its uncertainties, an upper limit for the
instanton cross section can be derived for any given choice of $r_I$.
For example, at $r_I= 10\%$, the upper limit would be 30 $\pb$, to be
compared with the theoretically predicted value of 8.9 $\pb$.
In a similar analysis by the H1 collaboration \citeHOneInst,
in a region of phase space with a lower range of $Q^2$
($10 \gev^2 < Q^2 < 100 \gev^2$), and at a comparable value $r_I=10\%$, an
 upper limit of
221 $\pb$ at a 95 \% confidence level (c.l.) was reported.
 This value is about a factor of five above the corresponding theoretical
 prediction. 

In order to derive a conservative upper limit without an explicit choice
of $r_I$,
the $t$ distribution was computed for a range of values of the instanton
 cross section. For a specific choice of the cross section, the 
$t$ distributions for data and instanton events become equal at
a certain specific value $t_0$, the instanton distribution overshooting the
 data
for $t > t_0$. An upper limit can then be set by choosing the instanton cross
section such that the number of instanton events exceeds the number
of data events for $t > t_0$ at a (one sided) 95\% c.l.
This method yields an upper limit at a 95\% c.l.
 for the instanton cross section, $\sigma_{\rm inst}$, of
\[ \sigma_{\rm inst} < 26 \pb \\ , \]
to be compared with the theoretically predicted cross section of $8.9 \pb$.
 The fraction of instanton-induced events remaining after the cut was
$r_I = 4.6\%$.

%
%
\section{Conclusion}
\label{sect-Concl}

A search for QCD-instanton-induced events has been performed in
neutral current deep inelastic $ep$ scattering, based on an
integrated luminosity of $38 \pb^{-1}$
in the kinematic range $Q^2 > 120\gev^2$, $x>10^{-3}$.
Cuts on the Fisher 
discriminant have been used to obtain instanton-enhanced subsamples.

Assuming that all data events belong to an instanton signal, a conservative
 background-independent  upper
 limit on the instanton cross section of $26 \pb$ at a 95\% c.l. has been set,
 to be compared
to the theoretically predicted cross section of 8.9 $\pb$.
%
%
\section{Acknowledgements}
\label{sect-Acknowldg}

We thank the DESY Directorate for their strong support and encouragement.
The HERA machine group and the DESY computing staff are gratefully
acknowledged for their outstanding operation of the collider and of the
data-analysis environment. We appreciate the contributions to the 
construction and maintenance of the ZEUS detector by many people who are
not listed as authors. It is a pleasure to thank F.~Schrempp for helpful
discussions.

%% file: zeus_defs.tex
\providecommand{\ZcoosysA}{%
The ZEUS coordinate system is a right-handed Cartesian system, with the $Z$
axis pointing in the proton beam direction, referred to as the ``forward
direction'', and the $X$ axis pointing left towards the center of HERA.
The coordinate origin is at the nominal interaction point.\xspace}
\providecommand{\ZcoosysB}{%
The ZEUS coordinate system is a right-handed Cartesian system, with the $Z$
axis pointing in the proton beam direction, referred to as the ``forward
direction'', and the $X$ axis pointing left towards the centre of HERA.
The coordinate origin is at the nominal interaction point.\xspace}
\providecommand{\Zpsrap}{%
The pseudorapidity is defined as $\eta=-\ln\left(\tan\frac{\theta}{2}\right)$,
where the polar angle, $\theta$, is measured with respect to the proton beam
direction.\xspace}
\providecommand{\ZcoosysfnA}{\footnote{\ZcoosysA}}
\providecommand{\ZcoosysfnB}{\footnote{\ZcoosysB}}
\providecommand{\ZcoosysfnAeta}{\footnote{\ZcoosysA\Zpsrap}}
\providecommand{\ZcoosysfnBeta}{\footnote{\ZcoosysB\Zpsrap}}
\providecommand{\Zdetdesc}{%
A detailed description of the ZEUS detector can be found 
elsewhere~\cite{zeus:1993:bluebook}. A brief outline of the 
components which are most relevant for this analysis is given
below.\xspace}
\providecommand{\Zctddesc}[1]{%
Charged particles are tracked in the central tracking detector (CTD)~\citeCTD,
which operates in a magnetic field of $1.43\Tesla$ provided by a thin 
superconducting coil. The CTD consists of 72~cylindrical drift chamber 
layers, organized in 9~superlayers covering the polar angle#1 region 
\mbox{$15^\circ<\theta<164^\circ$}. The transverse-momentum resolution for
full-length tracks is $\sigma(p_T)/p_T=0.0058p_T\oplus0.0065\oplus0.0014/p_T$,
with $p_T$ in $\Gev$.}
\providecommand{\Zcaldesc}{%
The high-resolution uranium--scintillator calorimeter (CAL)~\citeCAL consists 
of three parts: the forward (FCAL), the barrel (BCAL) and the rear (RCAL)
calorimeters. Each part is subdivided transversely into towers and
longitudinally into one electromagnetic section (EMC) and either one (in RCAL)
or two (in BCAL and FCAL) hadronic sections (HAC). The smallest subdivision of
the calorimeter is called a cell.  The CAL energy resolutions, as measured under
test beam conditions, are $\sigma(E)/E=0.18/\sqrt{E}$ for electrons and
$\sigma(E)/E=0.35/\sqrt{E}$ for hadrons ($E$ in $\Gev$).}
\providecommand{\xxx}{\hbox{$\qsm\qsm\qsm$}\xspace}
\providecommand{\DO}{{D{\O}}\xspace}
\providecommand{\Anti}[1]{\hbox{(anti-)}\discretionary{}{}{}#1}
\providecommand{\anti}[1]{(anti\discretionary{-)}{}{)}#1}
\chardef\usc=95
\chardef\til=126
\catcode`\@=11 
\DeclareRobustCommand\xdotspace{\futurelet\@let@token\@xdotspace}
\def\@xdotspace{%
  \ifx\@let@token.\else
  \ifx\@let@token\bgroup.\else
  \ifx\@let@token\egroup.\else
  \ifx\@let@token\/.\else
  \ifx\@let@token\ .\else
  \ifx\@let@token~.\else
  \ifx\@let@token!.\else
  \ifx\@let@token,.\else
  \ifx\@let@token:.\else
  \ifx\@let@token;.\else
  \ifx\@let@token?.\else
  \ifx\@let@token/.\else
  \ifx\@let@token'.\else
  \ifx\@let@token).\else
  \ifx\@let@token-.\else
  \ifx\@let@token\@xobeysp.\else
  \ifx\@let@token\space.\else
  \ifx\@let@token\@sptoken.\else
   .\space
   \fi\fi\fi\fi\fi\fi\fi\fi\fi\fi\fi\fi\fi\fi\fi\fi\fi\fi}
\catcode`\@=12 
\providecommand{\CL}[1]{$#1\%$~C.L\xdotspace}
\providecommand{\stru}[2]{%
   \relax\ifmmode\hbox{\vrule height#1 depth#2 width0pt}%
   \else\vrule height#1 depth#2 width0pt\fi}
\providecommand{\uline}[1]{$\underline{\hbox{#1\stru{0.pt}{.485ex}}}$}
\providecommand{\Ronum}[1]{\uppercase\expandafter{\romannumeral#1}}
\providecommand{\ronum}[1]{\expandafter{\romannumeral#1}}
\DeclareRobustCommand{\LaTeXZ}{%
  \LaTeX\kern-.05em4\kern-.1em
  {\raisebox{-0.2ex}{$\scriptstyle\mathrm{ZEUS}$}}\xspace}
\providecommand{\Auml}{\"A}
\providecommand{\Ouml}{\"O}
\providecommand{\Uuml}{\"U}
\providecommand{\auml}{\"a}
\providecommand{\ouml}{\"o}
\providecommand{\uuml}{\"u}
\providecommand{\eq}[1]{(\ref{eq-#1})}
\providecommand{\eqsand}[2]{(\ref{eq-#1}) and~(\ref{eq-#2})}
\providecommand{\eqsto}[2]{(\ref{eq-#1})--(\ref{eq-#2})}
\providecommand{\eqstwo}[2]{(\ref{eq-#1},\ref{eq-#2})}
\providecommand{\eqsthr}[3]{(\ref{eq-#1},\ref{eq-#2},\ref{eq-#3})}
\providecommand{\eqsfou}[4]{(\ref{eq-#1},\ref{eq-#2},\ref{eq-#3},\ref{eq-#4})}
\providecommand{\Eq}[1]{Equation~(\ref{eq-#1})}
\providecommand{\Eqsto}[2]{Equations~(\ref{eq-#1})--(\ref{eq-#2})}
\providecommand{\fig}[1]{Fig.~\ref{fig-#1}}
\providecommand{\Fig}[1]{Figure~\ref{fig-#1}}
\providecommand{\figand}[2]{Figs.~\ref{fig-#1} and~\ref{fig-#2}}
\providecommand{\Figand}[2]{Figures~\ref{fig-#1} and~\ref{fig-#2}}
\providecommand{\tab}[1]{Table~\ref{tab-#1}}
\providecommand{\Tab}[1]{Table~\ref{tab-#1}}
\providecommand{\taband}[2]{Tables~\ref{tab-#1} and~\ref{tab-#2}}
\providecommand{\cha}[1]{Chap.~\ref{cha-#1}}
\providecommand{\Cha}[1]{Chapter~\ref{cha-#1}}
\providecommand{\sect}[1]{Sect.~\ref{sec-#1}}
\providecommand{\Sect}[1]{Section~\ref{sec-#1}}
\providecommand{\sectand}[2]{Sects.~\ref{sec-#1} and~\ref{sec-#2}}
\providecommand{\sectto}[2]{Sects.~\ref{sec-#1} to~\ref{sec-#2}}
\providecommand{\Sectand}[2]{Sections~\ref{sec-#1} and~\ref{sec-#2}}
\providecommand{\Sectto}[2]{Sections~\ref{sec-#1} to~\ref{sec-#2}}
\providecommand{\LogMess}[1]{\typeout{^^J==> #1^^J}}
\providecommand{\eVdist}{\kern-0.06667em}
\providecommand{\Ev}{\ensuremath{\mathrm{e}\eVdist\mathrm{V\/}}}
\providecommand{\Kev}{\ensuremath{\mathrm{ke}\eVdist\mathrm{V\/}}}
\providecommand{\Mev}{\ensuremath{\mathrm{Me}\eVdist\mathrm{V\/}}}
\providecommand{\Gev}{\ensuremath{\mathrm{Ge}\eVdist\mathrm{V\/}}}
\providecommand{\Tev}{\ensuremath{\mathrm{Te}\eVdist\mathrm{V\/}}}
\providecommand{\ev}{\ensuremath{\,\mathrm{e}\eVdist\mathrm{V\/}}}
\providecommand{\kev}{\ensuremath{\,\mathrm{ke}\eVdist\mathrm{V\/}}}
\providecommand{\mev}{\ensuremath{\,\mathrm{Me}\eVdist\mathrm{V\/}}}
\providecommand{\gev}{\ensuremath{\,\mathrm{Ge}\eVdist\mathrm{V\/}}}
\providecommand{\tev}{\ensuremath{\,\mathrm{Te}\eVdist\mathrm{V\/}}}
\providecommand{\mb}{\ensuremath{\,\mathrm{mb}}}
\providecommand{\mub}{\ensuremath{\,\upmu\mathrm{b}}}
\providecommand{\nb}{\ensuremath{\,\mathrm{nb}}}
\providecommand{\pb}{\ensuremath{\,\mathrm{pb}}}
\providecommand{\fb}{\ensuremath{\,\mathrm{fb}}}
\providecommand{\mubi}{\ensuremath{\,\upmu\mathrm{b}^{-1}}}
\providecommand{\mbi}{\ensuremath{\,\mathrm{mb}^{-1}}}
\providecommand{\nbi}{\ensuremath{\,\mathrm{nb}^{-1}}}
\providecommand{\pbi}{\ensuremath{\,\mathrm{pb}^{-1}}}
\providecommand{\fbi}{\ensuremath{\,\mathrm{fb}^{-1}}}
\providecommand{\met}{\ensuremath{\,\mathrm{m}}}
\providecommand{\nm}{\ensuremath{\,\mathrm{nm}}}
\providecommand{\mum}{\ensuremath{\,\upmu\mathrm{m}}}
\providecommand{\mm}{\ensuremath{\,\mathrm{mm}}}
\providecommand{\cm}{\ensuremath{\,\mathrm{cm}}}
\providecommand{\km}{\ensuremath{\,\mathrm{km}}}
\providecommand{\Amp}{\ensuremath{\,\mathrm{A}}}
\providecommand{\nA}{\ensuremath{\,\mathrm{nA}}}
\providecommand{\muA}{\ensuremath{\,\upmu\mathrm{A}}}
\providecommand{\mA}{\ensuremath{\,\mathrm{mA}}}
\providecommand{\kA}{\ensuremath{\,\mathrm{kA}}}
\providecommand{\Hz}{\ensuremath{\,\mathrm{Hz}}}
\providecommand{\kHz}{\ensuremath{\,\mathrm{kHz}}}
\providecommand{\MHz}{\ensuremath{\,\mathrm{MHz}}}
\providecommand{\GHz}{\ensuremath{\,\mathrm{GHz}}}
\providecommand{\ppm}{\ensuremath{\,\mathrm{ppm}}}
\providecommand{\ppb}{\ensuremath{\,\mathrm{ppb}}}
\providecommand{\scd}{\ensuremath{\,\mathrm{s}}}
\providecommand{\ps}{\ensuremath{\,\mathrm{ps}}}
\providecommand{\ns}{\ensuremath{\,\mathrm{ns}}}
\providecommand{\mus}{\ensuremath{\,\upmu\mathrm{s}}}
\providecommand{\ms}{\ensuremath{\,\mathrm{ns}}}
\providecommand{\rad}{\ensuremath{\,\mathrm{rad}}}
\providecommand{\murad}{\ensuremath{\,\upmu\mathrm{rad}}}
\providecommand{\mrad}{\ensuremath{\,\mathrm{mrad}}}
\providecommand{\gra}{\ensuremath{\,\mathrm{g}}}
\providecommand{\pg}{\ensuremath{\,\mathrm{pg}}}
\providecommand{\nang}{\ensuremath{\,\mathrm{ng}}}
\providecommand{\mug}{\ensuremath{\,\upmu\mathrm{g}}}
\providecommand{\mg}{\ensuremath{\,\mathrm{mg}}}
\providecommand{\kg}{\ensuremath{\,\mathrm{kg}}}
\providecommand{\Tesla}{\ensuremath{\,\mathrm{T}}}
\providecommand{\Kelvin}{\ensuremath{\,\mathrm{K}}}
\providecommand{\odiv}{\ensuremath{\operatorname*{div}}}
\providecommand{\ograd}{\ensuremath{\operatorname*{grad}}}
\providecommand{\orot}{\ensuremath{\operatorname*{rot}}}
\providecommand{\odiag}{\ensuremath{\operatorname*{diag}}}
\providecommand{\ocov}{\ensuremath{\operatorname*{cov}}}
\providecommand{\slashfrac}[2]{%
  \raisebox{0.5ex}{\ensuremath #1}\kern-0.12em/\kern-0.08em
  \raisebox{-.8ex}{\ensuremath #2}}
\providecommand{\intl}{\ensuremath{\int\limits}}
\providecommand{\ointl}{\ensuremath{\oint\limits}}
\providecommand{\sqr}[3]{%
    {\vcenter{\hrule height.#3ex\hbox{\vrule width.#2ex height#1ex
     \kern#1ex\vrule width.#3ex}\hrule height.#2ex}}}
\providecommand{\dalem}{\mathchoice\sqr{1.3}{08}{24}\sqr{1.4}{08}{24}
                    \sqr{1.15}{07}{21}\sqr{1.0}{06}{18}\,}
\providecommand{\vect}[1]{\ensuremath{\begin{matrix}#1\end{matrix}}}
\providecommand{\pvect}[1]{\ensuremath{\begin{pmatrix}#1\end{pmatrix}}}
\providecommand{\bvect}[1]{\ensuremath{\begin{bmatrix}#1\end{bmatrix}}}
\providecommand{\vvect}[1]{\ensuremath{\begin{vmatrix}#1\end{vmatrix}}}
\providecommand{\Vvect}[1]{\ensuremath{\begin{Vmatrix}#1\end{Vmatrix}}}
\providecommand{\widebar}[1]{%
   \mkern1.5mu\overline{\mkern-1.5mu#1\mkern-1.mu}\mkern1.mu}
\catcode`\@=11 
\providecommand{\parenbar}{\ensuremath{\mathpalette\p@renb@r}}
\def\p@renb@r#1#2{\vbox{%
  \ifx#1\scriptscriptstyle \dimen@.7em\dimen@ii.2em\else
  \ifx#1\scriptstyle \dimen@.8em\dimen@ii.25em\else
  \dimen@1em\dimen@ii.4em\fi\fi \offinterlineskip
  \ialign{\hfill##\hfill\cr
    \vbox{\hrule width\dimen@ii}\cr
    \noalign{\vskip-.3ex}%
    \hbox to\dimen@{$\mathchar300\hfil\mathchar301$}\cr
    \noalign{\vskip-.3ex}%
    $#1#2$\cr}}}
\catcode`\@=12 
\providecommand{\nuan}{\parenbar{\nu}}
\providecommand{\nunubar}{\parenbar{\nu}}
\providecommand{\ppbar}{\parenbar{p}}
\providecommand{\qqbar}{\parenbar{q}}
\providecommand{\pbar}{\widebar{p}}
\providecommand{\nbar}{\widebar{n}}
\providecommand{\qbar}{\widebar{q}}
\providecommand{\dbar}{\widebar{d}}
\providecommand{\ubar}{\widebar{u}}
\providecommand{\sbar}{\widebar{s}}
\providecommand{\cbar}{\widebar{c}}
\providecommand{\bbar}{\widebar{b}}
\providecommand{\tbar}{\widebar{t}}
\providecommand{\nubar}{\widebar{\nu}}
\providecommand{\Dbar}{\widebar{D}}
\providecommand{\ebar}{\widebar{e}}
\providecommand{\Ebar}{\widebar{e}}
\providecommand{\Hebar}{\widebar{\mathrm{He}}}
\providecommand{\Cbar}{\widebar{\mathrm{C}}}
\providecommand{\Abar}{\widebar{\mathrm{A}}}
\providecommand{\Kbar}{\widebar{K}}
\providecommand{\chibar}{\widebar{\chi}}
\providecommand{\LQbar}{\widebar{\rm LQ}}
\providecommand{\zero}{{\scriptscriptstyle 0}}
\providecommand{\sone}{{\scriptscriptstyle 1}}
\providecommand{\stwo}{{\scriptscriptstyle 2}}
\providecommand{\sthr}{{\scriptscriptstyle 3}}
\providecommand{\sfou}{{\scriptscriptstyle 4}}
\providecommand{\sfiv}{{\scriptscriptstyle 5}}
\providecommand{\ssix}{{\scriptscriptstyle 6}}
\providecommand{\ssev}{{\scriptscriptstyle 7}}
\providecommand{\seig}{{\scriptscriptstyle 8}}
\providecommand{\snin}{{\scriptscriptstyle 9}}
\providecommand{\sten}{{\scriptscriptstyle 10}}
\providecommand{\half}{{\scriptscriptstyle 1/2}}
\providecommand{\MSbar}{\ensuremath{\hbox{$\overline{\rm MS}$}\xspace}}
\providecommand{\UO}{{\ensuremath{\rm U}(1)_\sY}}
\providecommand{\SUT}{\ensuremath{{\rm SU}(2)_\sL}}
\providecommand{\SUTH}{\ensuremath{{\rm SU}(3)_c}}
\providecommand{\SUTUO}{\ensuremath{{\rm SU}(2)_\sL\times{\rm U}(1)_\sY}}
\providecommand{\SUTSUTUO}{\ensuremath{{\rm SU}(3)_c\times{\rm SU}(2)_\sL\times{\rm U}(1)_\sY}}
\providecommand{\alsmu}[1]{\ensuremath{{\alpha_s(\mu_{#1}^2)}}}
\providecommand{\alsmz}{\ensuremath{{\alpha_s(M_Z^2)}}}
\providecommand{\alsqs}{\ensuremath{{\alpha_s(Q^2)}}}
\providecommand{\als}{\ensuremath{\alpha_s}}
\providecommand{\ctwb}{\ensuremath{\cos\theta_W}}
\providecommand{\ctws}{\ensuremath{\cos^2\theta_W}}
\providecommand{\diff}{\ensuremath{{\rm d}}}
\providecommand{\gh}{\ensuremath{{\gamma_h}}}
\providecommand{\lsq}[1]{\ensuremath{{\lambda_{#1}}}}
\providecommand{\lsqp}[1]{\ensuremath{{\lambda'_{#1}}}}
\providecommand{\lsqpp}[1]{\ensuremath{{{\lambda''}_{#1}}}}
\providecommand{\rnge}{\ensuremath{\hbox{$\,\mathrm{--}\,$}}}
\providecommand{\sihat}{\ensuremath{{\hat\sigma}}}
\providecommand{\sitil}{\ensuremath{{\tilde\sigma}}}
\providecommand{\stwb}{\ensuremath{\sin\theta_W}}
\providecommand{\stwss}{\ensuremath{\sin^4\theta_W}}
\providecommand{\stws}{\ensuremath{\sin^2\theta_W}}
\providecommand{\ta}{{\ensuremath{\theta^\ast}}}
\providecommand{\tilR}{\ensuremath{\tilde R}}
\providecommand{\tilS}{\ensuremath{\tilde S}}
\providecommand{\tilU}{\ensuremath{\tilde U}}
\providecommand{\tilV}{\ensuremath{\tilde V}}
\providecommand{\tl}{\ensuremath{{\theta_\ell}}}
\providecommand{\tw}{\ensuremath{\theta_W}}
\providecommand{\Born}{\ensuremath{{\rm Born}}}
\providecommand{\BR}{\ensuremath{{\rm BR}}}
\providecommand{\CC}{\ensuremath{{\rm CC}}}
\providecommand{\CI}{\ensuremath{{\rm CI}}}
\providecommand{\CJ}{\ensuremath{{\rm CJ}}}
\providecommand{\DA}{\ensuremath{{\rm DA}}}
\providecommand{\DCA}{\ensuremath{{\rm DCA}}}
\providecommand{\GRV}{{\ensuremath{\rm GRV}}}
\providecommand{\IC}{\ensuremath{{\rm IC}}}
\providecommand{\ISR}{\ensuremath{{\rm ISR}}}
\providecommand{\IP}{\ensuremath{{\rm I$\kern-0.01667em$P}\xspace}}
\providecommand{\JB}{\ensuremath{{\rm JB}}}
\providecommand{\LL}{\ensuremath{{\rm LL}}}
\providecommand{\LQ}{\ensuremath{{\rm LQ}}}
\providecommand{\LR}{\ensuremath{{\rm LR}}}
\providecommand{\MC}{\ensuremath{{\rm MC}}}
\providecommand{\NC}{\ensuremath{{\rm NC}}}
\providecommand{\NS}{\ensuremath{{\rm NS}}}
\providecommand{\QPM}{\ensuremath{{\rm QPM}}}
\providecommand{\RL}{\ensuremath{{\rm RL}}}
\providecommand{\RPv}{\ensuremath{{{\not R}_P}}}
\providecommand{\RR}{\ensuremath{{\rm RR}}}
\providecommand{\SM}{\ensuremath{{\rm SM}}}
\providecommand{\bkg}{\ensuremath{{\rm bkg}}}
\providecommand{\cut}{\ensuremath{{\rm cut}}}
\providecommand{\data}{{\ensuremath{\rm data}}}
\providecommand{\sdet}{\ensuremath{{\rm det}}}
\providecommand{\dof}{\ensuremath{{\rm dof}}}
\providecommand{\eff}{\ensuremath{{\rm eff}}}
\providecommand{\elm}{\ensuremath{{\rm elm}}}
\providecommand{\evt}{\ensuremath{{\rm evt}}}
\providecommand{\exc}{\ensuremath{{\rm exc}}}
\providecommand{\fit}{\ensuremath{{\rm fit}}}
\providecommand{\had}{\ensuremath{{\rm had}}}
\providecommand{\jet}{\ensuremath{{\rm jet}}}
\providecommand{\meas}{\ensuremath{{\rm meas}}}
\providecommand{\miss}{\ensuremath{{\rm miss}}}
\providecommand{\obs}{\ensuremath{{\rm obs}}}
\providecommand{\pnt}{{\ensuremath{\rm pnt}}}
\providecommand{\sL}{\ensuremath{{\rm L}}}
\providecommand{\sR}{\ensuremath{{\rm R}}}
\providecommand{\sS}{\ensuremath{{\rm S}}}
\providecommand{\sT}{\ensuremath{{\rm T}}}
\providecommand{\sY}{\ensuremath{{\rm Y}}}
\providecommand{\sca}{\ensuremath{{\rm sca}}}
\providecommand{\sexp}{\ensuremath{{\rm exp}}}
\providecommand{\sint}{\ensuremath{{\rm int}}}
\providecommand{\smin}{\ensuremath{{\rm min}}}
\providecommand{\smax}{\ensuremath{{\rm max}}}
\providecommand{\srad}{\ensuremath{{\rm rad}}}
\providecommand{\sthe}{\ensuremath{{\rm the}}}
\providecommand{\sys}{\ensuremath{{\rm sys}}}
\providecommand{\tot}{\ensuremath{{\rm tot}}}
\providecommand{\true}{\ensuremath{{\rm true}}}
\providecommand{\Cor}{\ensuremath{{\cal C}}}
\providecommand{\F}{{\ensuremath{\cal F}}}
\providecommand{\Lumi}{\ensuremath{{\cal L}}}
\providecommand{\Prob}{\ensuremath{{\cal P}}}
\providecommand{\ord}[1]{\ensuremath{{\cal O}(#1)}}
\providecommand{\ordlr}[1]{\ensuremath{{\cal O}\left(#1\right)}}
\mathchardef\qsm=63
\mathchardef\pls=43
\mathchardef\mns=512
\mathchardef\plm=518
\mathchardef\eql=61
\mathchardef\smallleft=300
\mathchardef\smallright=301
\mathchardef\les=316
\mathchardef\gre=318
\mathchardef\leq=532
\mathchardef\grq=533
\providecommand{\cA}{{\phantom{0}}}
\providecommand{\cB}{{\phantom{00}}}
\providecommand{\cC}{{\phantom{000}}}
\providecommand{\cD}{{\phantom{0000}}}
\providecommand{\cAp}{{\phantom{0.}}}
\providecommand{\cM}{{\phantom{\mns}}}
\providecommand{\cP}{{\phantom{\pls}}}
\catcode`\@=11 
\setlength{\pict@scale}{0.1mm}
\providecommand{\psfigadd}[4]{%
\setcounter{pict@width}{1*\ratio{#2+\pict@scale/2}{\pict@scale}}
\setcounter{pict@height}{1*\ratio{#3+\pict@scale/2}{\pict@scale}}
\setlength{\unitlength}{\pict@scale}
\hbox to #2{\hspace{-\fill}\begin{picture}(\thepict@width,\thepict@height)
\put(0,0){\psfig{figure=#1,width=#2,height=#3,clip=}}
\SetScale{0.283466457}
\SetWidth{1.763889}
{#4}
\end{picture}}
}
\providecommand{\psfigaddtwo}[7]{%
\setcounter{pict@widthfst}{1*\ratio{#2+\pict@scale/2}{\pict@scale}}
\setcounter{pict@widthscd}{1*\ratio{#2+#4+\pict@scale/2}{\pict@scale}}
\setcounter{pict@widthtot}{1*\ratio{#2+#4+#6+\pict@scale/2}{\pict@scale}}
\setcounter{pict@height}{1*\ratio{#3+\pict@scale/2}{\pict@scale}}
\setlength{\unitlength}{\pict@scale}
\hbox{\hspace{-\fill}\begin{picture}(\thepict@widthtot,\thepict@height)
\put(0,0){\psfig{figure=#1,width=#2,height=#3,clip=}}
\put(\thepict@widthscd,0){\psfig{figure=#5,width=#6,height=#3,clip=}}
\SetScale{0.283466457}
\SetWidth{1.763889}
{#7}
\end{picture}}
}
\providecommand{\psfigror}[4]{%
\setcounter{pict@width}{1*\ratio{#2+\pict@scale/2}{\pict@scale}}
\setcounter{pict@height}{1*\ratio{#3+\pict@scale/2}{\pict@scale}}
\setlength{\unitlength}{\pict@scale}
\hbox{\begin{picture}(\thepict@width,\thepict@height)
\put(0,\thepict@height){\psfig{figure=#1,width=#3,height=#2,clip=,angle=270}}
\SetScale{0.283466457}
\SetWidth{1.763889}
{#4}
\end{picture}}
}
\providecommand{\psfigrol}[4]{%
\setcounter{pict@width}{1*\ratio{#2+\pict@scale/2}{\pict@scale}}
\setcounter{pict@height}{1*\ratio{#3+\pict@scale/2}{\pict@scale}}
\setlength{\unitlength}{\pict@scale}
\hbox{\begin{picture}(\thepict@width,\thepict@height)
\put(0,0){\psfig{figure=#1,width=#3,height=#2,clip=,angle=90}}
\SetScale{0.283466457}
\SetWidth{1.763889}
{#4}
\end{picture}}
}
\catcode`\@=12 
\providecommand{\shiftfloat}[1]{
  \setlength{\listtextwidth}{\textwidth}
  \addtolength{\listtextwidth}{-#1}}
\providecommand{\none}{\hbox{---}}
\providecommand{\pcite}[1]{{\protect\cite{#1}}}
\providecommand{\pnl}{\protect{\newline}}
\catcode`\@=11 
\providecommand{\tabfootnote}[2]{%
  \setlength{\@tabfninsert}{0.8em}
  \setlength{\@tabfnwidth}{\textwidth}
  \addtolength{\@tabfnwidth}{-\@tabfninsert}
  \addtolength{\@tabfnwidth}{-0.4em}
  \noindent\makebox[\@tabfninsert][r]{\footnotesize$^{#1}$\hfil}\hfill%
  \parbox[t]{\@tabfnwidth}{\footnotesize #2\hfill}}
\catcode`\@=12 

%% file: defs.tex
\setcounter{FiguresAvailable}{1}
\setlength{\halftextwidth}{0.48\textwidth}
\setlength{\pictwidth}{1.\textwidth}
\providecommand{\PlotDir}{epsfiles}
\providecommand{\CondFigEW}[5]{%
%
\ifthenelse{\value{FiguresAvailable} = 1}{
\begin{figure}[h!]
\begin{center}
 \mbox{\epsfig{figure=#4,width=#5\textwidth}}
\caption{#3}
\label{fig-#1}
\end{center}
\end{figure}}%
{
\begin{figure}[h]
\caption{}
\label{fig-#1}
\end{figure}
\vspace*{-1.5ex}
\begin{center}
 \mbox{\textbf{eps file #4 to be inserted here}}
\end{center}
\vspace*{1.ex}
}}
\providecommand{\CondFigE}[4]{%
%
\ifthenelse{\value{FiguresAvailable} = 1}{
\begin{figure}[h!]
\begin{center}
 \mbox{\epsfig{figure=#4,width=\textwidth}}
\caption{#3}
\label{fig-#1}
\end{center}
\end{figure}}%
{
\begin{figure}[h]
\caption{}
\label{fig-#1}
\end{figure}
\vspace*{-1.5ex}
\begin{center}
 \mbox{\textbf{eps file #4 to be inserted here}}
\end{center}
\vspace*{1.ex}
}}
\providecommand{\CondFigZ}[5]{%
%
\ifthenelse{\value{FiguresAvailable} = 1}{
\begin{figure}[h!]
\begin{center}
\begin{tabular}{llll}
 \hspace*{2.em}\hspace*{-\halftextwidth}\mbox{\epsfig{figure=#4,
      width=0.49\textwidth}} & 
 \hspace*{-\halftextwidth} a)\hspace*{-1.0em}\hspace*{\halftextwidth} &
 \hspace*{-1.0em}\mbox{\epsfig{figure=#5,
      width=0.49\textwidth}} &
 \hspace*{-\halftextwidth} b) \\
 \end{tabular}  
%
\caption{#3}
\label{fig-#1}
\end{center}
\end{figure}}%
{
\begin{figure}[h]
\caption{}
\label{fig-#1}
\end{figure}
\vspace*{-2.ex}
\begin{center}
\textbf{eps files\\ 
\begin{tabular}{cc}
 #4 & #5
\end{tabular}\\
to be inserted here}
\end{center}
\vspace*{1.ex}
}}
\providecommand{\CondFigD}[6]{%
%
\ifthenelse{\value{FiguresAvailable} = 1}{
\begin{figure}[h!]
\begin{center}
\begin{tabular}{lclc}
 \hspace*{2.em}\hspace*{-\halftextwidth}\mbox{\epsfig{figure=#4,
      width=0.49\textwidth}} &
 \hspace*{-\halftextwidth} a)\hspace*{-1.0em}\hspace*{\halftextwidth} &
 \hspace*{-1.0em}\mbox{\epsfig{figure=#5,
      width=0.49\textwidth}} &
 \hspace*{-\halftextwidth} b) \\[.5ex]
 \hspace*{2.em}\hspace*{-\halftextwidth}\mbox{\epsfig{figure=#6,
      width=0.49\textwidth}} &
 \hspace*{-\halftextwidth} c)\hspace*{-1.0em}\hspace*{\halftextwidth} & \\
\end{tabular}  
%
\caption{#3}
\label{fig-#1}
\end{center}
\end{figure}}%
{
\begin{figure}[h]
\caption{}
\label{fig-#1}
\end{figure}
\vspace*{-2.ex}
\begin{center}
 \textbf{eps files\\ 
\begin{tabular}{cc}
 #4 & #5\\
 #6 &
\end{tabular}\\
to be inserted here}
\end{center}
\vspace*{1.ex}
}}
\providecommand{\CondFigRowD}[9]{%
%
\ifthenelse{\value{FiguresAvailable} = 1}{
\begin{figure}[h!]
\begin{center}
\begin{tabular}{ccc}
 \mbox{\epsfig{figure=#4,width=#7\textwidth}} &
 \mbox{\epsfig{figure=#5,width=#8\textwidth}} &
 \mbox{\epsfig{figure=#6,width=#9\textwidth}}
\end{tabular}\\
\begin{tabular}{lll}
 \hspace*{#7\halftextwidth} $\mathrm{(a)}$ \hspace*{#7\halftextwidth}&
 \hspace*{#8\halftextwidth} $\mathrm{(b)}$ \hspace*{#8\halftextwidth}&
 \hspace*{#9\halftextwidth} $\mathrm{(c)}$ \hspace*{#9\halftextwidth} \\
\end{tabular}
%
\caption{#3}
\label{fig-#1}
\end{center}
\end{figure}}%
{
\begin{figure}[h]
\caption{}
\label{fig-#1}
\end{figure}
\vspace*{-2.ex}
\begin{center}
 \textbf{eps files\\ 
\begin{tabular}{cc}
 #4 & #5\\
 #6 &
\end{tabular}\\
to be inserted here}
\end{center}
\vspace*{1.ex}
}}
\providecommand{\CondFigV}[7]{%
%
\ifthenelse{\value{FiguresAvailable} = 1}{
\begin{figure}[h!]
\begin{center}
\begin{tabular}{lclc}
 \hspace*{2.em}\hspace*{-\halftextwidth}\mbox{\epsfig{figure=#4,
      width=0.49\textwidth}} &
 \hspace*{-\halftextwidth} a)\hspace*{-1.0em}\hspace*{\halftextwidth} &
 \hspace*{-1.0em}\mbox{\epsfig{figure=#5,
      width=0.49\textwidth}} &
 \hspace*{-\halftextwidth} b) \\[.5ex]
 \hspace*{2.em}\hspace*{-\halftextwidth}\mbox{\epsfig{figure=#6,
      width=0.49\textwidth}} &
 \hspace*{-\halftextwidth} c)\hspace*{-1.0em}\hspace*{\halftextwidth} &
 \hspace*{-1.0em}\mbox{\epsfig{figure=#7,
      width=0.49\textwidth}} &
 \hspace*{-\halftextwidth} d) \\
\end{tabular}  
%
\caption{#3}
\label{fig-#1}
\end{center}
\end{figure}}%
{
\begin{figure}[h]
\caption{}
\label{fig-#1}
\end{figure}
\vspace*{-2.ex}
\begin{center}
 \textbf{eps files\\ 
\begin{tabular}{cc}
 #4 & #5\\
 #6 & #7
\end{tabular}\\
to be inserted here}
\end{center}
\vspace*{1.ex}
}}
%
\providecommand{\ptmiss}{\ensuremath{\slash\hspace*{-0.7em}{p_t}}}
\providecommand{\scalEt}{\ensuremath{E_{\mathrm{t,s}}}}
\providecommand{\Qprisqd}{\ensuremath{{Q'}^{\hspace*{0.1em}2}}}
\providecommand{\QprisqdDA}{\ensuremath{{Q'}_\DA^{\hspace*{0.2em}2}}}
\providecommand{\Qprisqdtr}{\ensuremath{{Q'}_{\mathrm{true}}^{\hspace*{0.2em}2}}}
\providecommand{\Qprisqdmin}{\ensuremath{{Q'}_{\mathrm{min}}^{\hspace*{0.2em}2}}}
\providecommand{\regmeth}{region method }
\providecommand{\RegMeth}{Region Method }
\providecommand{\gen}{{\rm gen}}
\providecommand{\Hwg}{{\rm Hwg}}
\providecommand{\Dja}{{\rm Dja}}
\providecommand{\EtaInst}{\ensuremath{\eta_{\mathrm{Inst}}}}
\providecommand{\MInst}{\ensuremath{M_{\mathrm{Inst}}}}
\providecommand{\lPtJet}{\ensuremath{\mathrm{log}_{\mathrm{10}}\ p_T^{\mathrm{Jet}}}}
\providecommand{\lPtInst}{\ensuremath{\mathrm{log}_{\mathrm{10}}\ P_t(\mathrm{Inst})}}
\providecommand{\GammaHad}{\ensuremath{\gamma_{\mathrm{h}}}}
\providecommand{\EpsP}{\ensuremath{{\epsilon'}}}
\providecommand{\Circ}{\ensuremath{C}}
\providecommand{\Sph}{\ensuremath{S}}
\providecommand{\NzufosI}{\ensuremath{N_{\mathrm{efo}}}}
\providecommand{\Nzuftr}{\ensuremath{N_{\mathrm{eft}}}}
\providecommand{\ptmiss}{\ensuremath{\ \slash\hspace*{-0.7em}{p_t}}}
\providecommand{\Ets}{\ensuremath{E_{\mathrm{t,s}}}}

%% file: InstPaper-ref.tex
{
\def\bibname{\Large\bf References}
\def\refname{\Large\bf References}
\pagestyle{plain}
\ifzeusbst
  \bibliographystyle{./BiBTeX/bst/l4z_default}
\fi
\ifzdrftbst
  \bibliographystyle{./BiBTeX/bst/l4z_draft}
\fi
\ifzbstepj
  \bibliographystyle{./BiBTeX/bst/l4z_epj}
\fi
\ifzbstnp
  \bibliographystyle{./BiBTeX/bst/l4z_np}
\fi
\ifzbstpl
  \bibliographystyle{./BiBTeX/bst/l4z_pl}
\fi
{\raggedright
\bibliography{./BiBTeX/user/syn.bib,%
              ./BiBTeX/bib/l4z_articles.bib,%
              ./BiBTeX/bib/l4z_books.bib,%
              ./BiBTeX/bib/l4z_conferences.bib,%
              ./BiBTeX/bib/l4z_h1.bib,%
              ./BiBTeX/bib/l4z_misc.bib,%
              ./BiBTeX/bib/l4z_old.bib,%
              ./BiBTeX/bib/l4z_preprints.bib,%
              ./BiBTeX/bib/l4z_replaced.bib,%
              ./BiBTeX/bib/l4z_temporary.bib,%
              ./BiBTeX/bib/l4z_zeus.bib}}
}
\vfill\eject

%% file: InstPaper-tab.tex
\setlength{\textheight}{30cm}
\appendix
\setcounter{table}{0}
\setlength{\textheight}{22.5cm}
%
%
\begin{table}
\begin{center}
\begin{tabular}{||c|c||c||c||c|c||c|c||}
\hline
 & $r_I$ [$\%$] & DATA & QCDINS & DJANGOH & $P_S$ & HERWIG & $P_S$ \\
\hline\hline
%
$t >  8.0$ & $32.6$ & $1847 \pm 43$   & $188.5 \pm 1.7$ & $2592   \pm 26$  & 12 & $2145   \pm 27$   & 14 \\
$t >  8.5$ & $24.0$ & $ 925 \pm 30$   & $139.0 \pm 1.4$ & $1338   \pm 19$  & 17 & $1091   \pm 19$   & 21 \\
$t >  9.0$ & $16.4$ & $ 424 \pm 21$   & $ 95.1 \pm 1.2$ & $ 630.2 \pm 13$  & 24 & $ 524.1 \pm 13$   & 29 \\
$t >  9.5$ & $10.1$ & $ 179 \pm 13$   & $ 58.4 \pm 0.9$ & $ 263.8 \pm 8.3$ & 36 & $ 229.5 \pm  8.8$ & 41 \\
$t > 10.0$ & $5.5$  & $  76 \pm  8.7$ & $ 31.8 \pm 0.7$ & $ 105.6 \pm 5.3$ & 49 & $  89.8 \pm  5.5$ & 58 \\
$t > 10.5$ & $2.7$  & $  33 \pm  5.7$ & $ 15.7 \pm 0.5$ & $  35.1 \pm 3.0$ & 73 & $  35.1 \pm  3.4$ & 73 \\
%
%
%
%
%
%
%
\hline
\end{tabular}
\end{center}
\caption{Numbers of events within instanton enhancing cuts chosen
such that a fraction $r_I$ of the QCDINS sample within 
fiducial cuts (see Sect.~\ref{subs-NCDISsel}) is kept.
Statistical errors are given. The separation power $P_S$ is also shown
 (see Sect.~\ref{sect-MethIE}). }
\label{tab-CutResults}
\end{table}\hspace*{-0.37em}

%% file: InstPaper-fig.tex
\input{zeus_defs}
\input{defs}
\CondFigEW{InstIndEvtKine}
{Kinematics of instanton-induced events at HERA}
{Kinematics of instanton-induced ep collisions.} 
{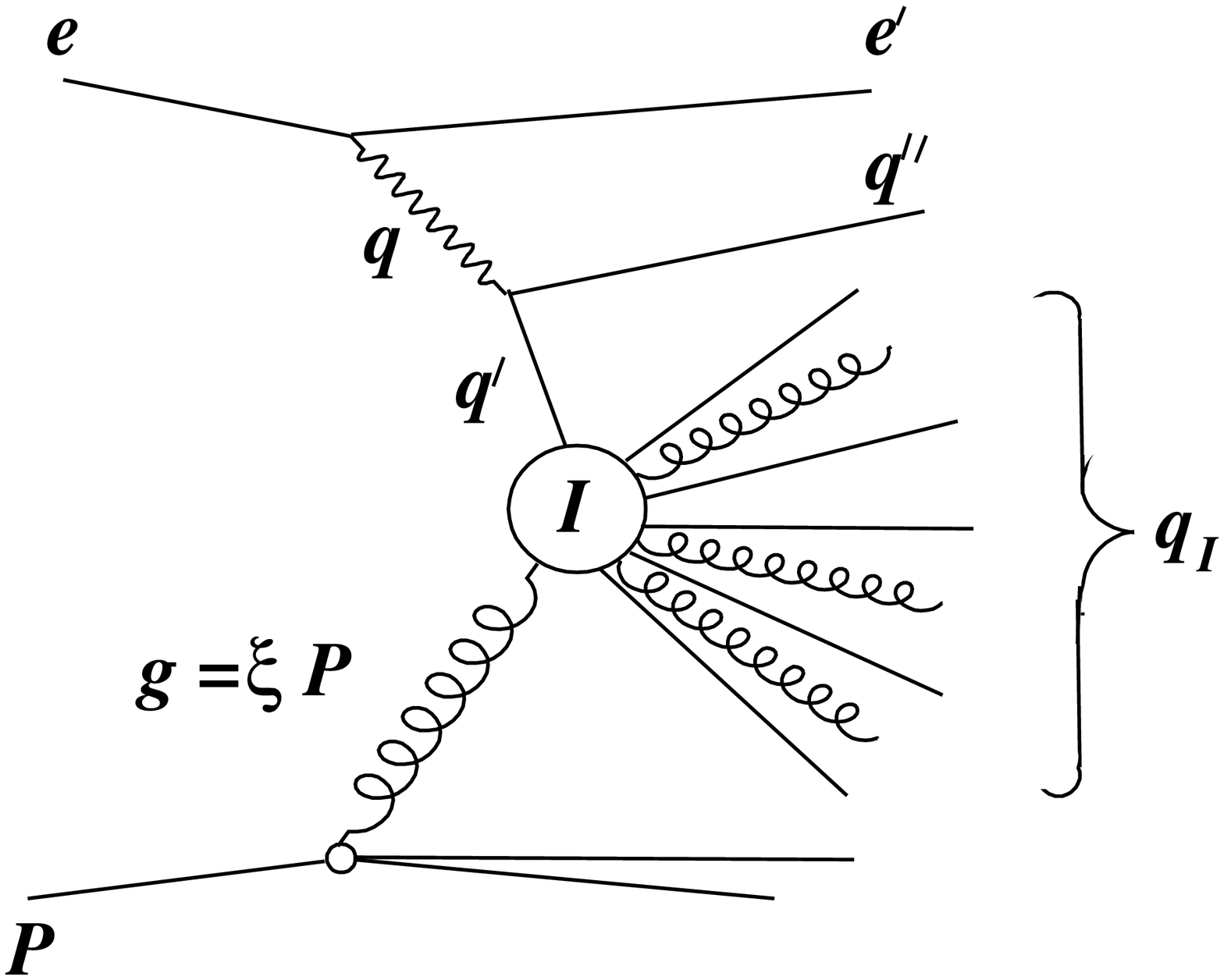,clip=}
{.7}
%
\CondFigEW{QPR} 
{The kinematic variable $\QprisqdDA$.}
{Distribution of the kinematic variable $\QprisqdDA$ describing the 
hard subprocess of instanton-induced events. Statistical error bars are given.}
{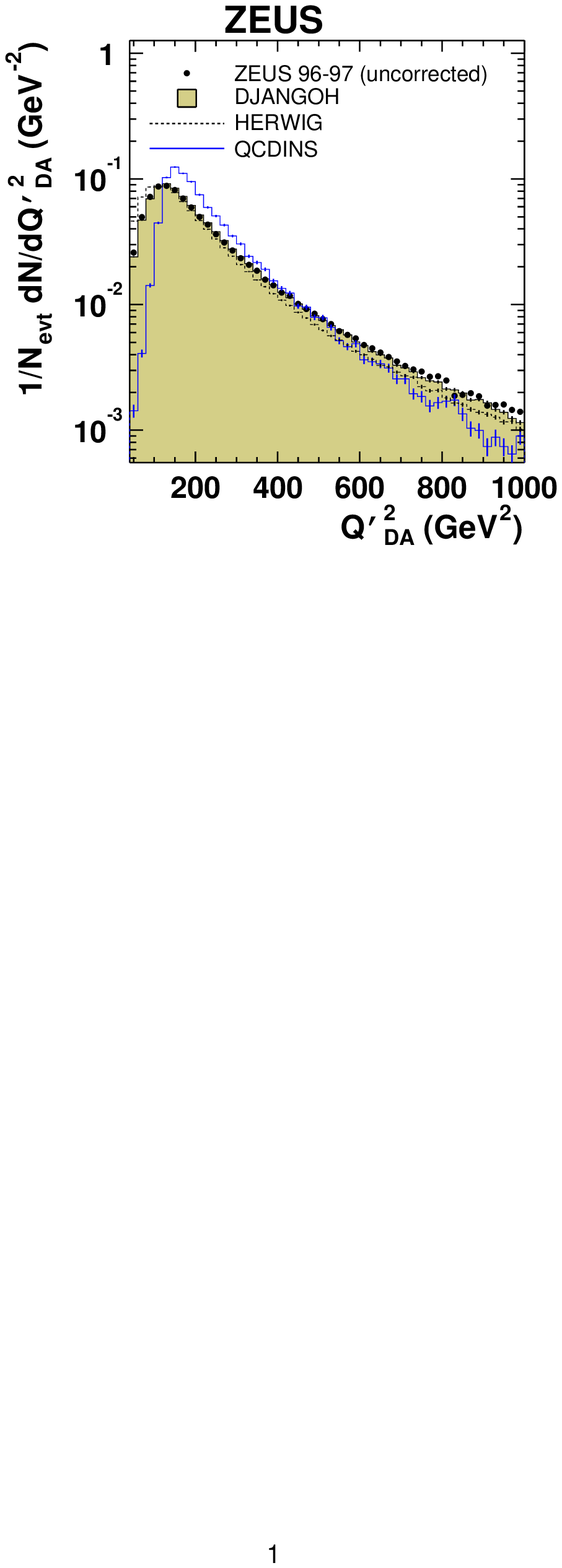,clip=} 
{0.5}
%
%
%
\CondFigEW{IEvarsOne} 
{Instanton candidate control plots: $\PtJet$, $\NzufosI$, $\Nzuftr$, $C$}
{Distributions of variables calculated from 
the instanton region, shown for the fiducial sample:
(a) the transverse momentum, $p_T^{\mathrm{jet}}$, of the current jet,
(b) EFO multiplicity, $\NzufosI$,
(c) EFO track multiplicity, $\Nzuftr$, 
(d) circularity, $C$, in the hcms. Statistical errors are given.}
{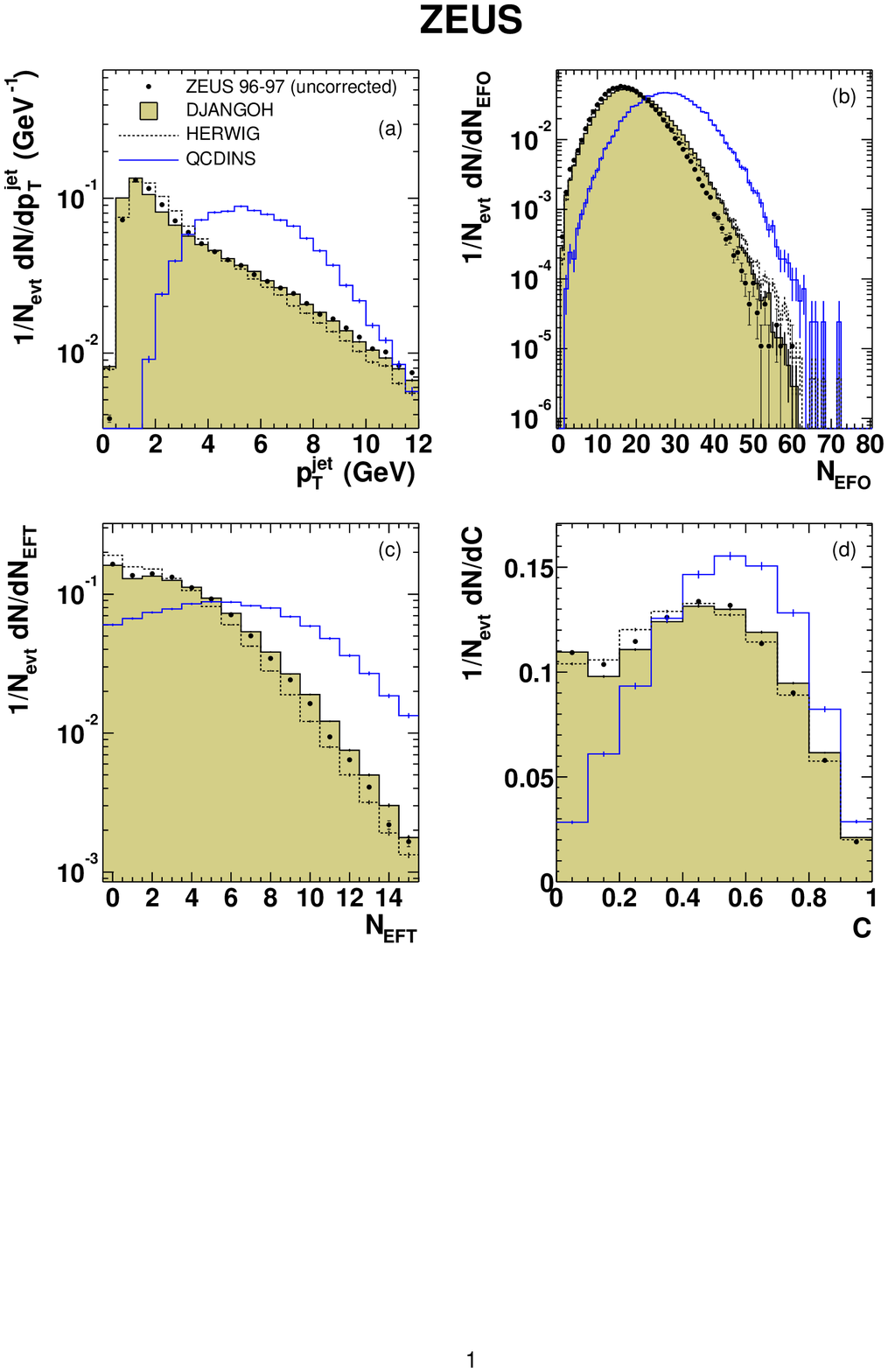,clip=} 
{1.0}
%
%
%
\CondFigEW{ShapeVars} 
{Shape variables $S_{\mathrm{det}}$ and $\EpsP$.}
{Distributions of two of the shape variables for the fiducial sample
%
(a) sphericity $S$,
(b) $\EpsP$.
}
{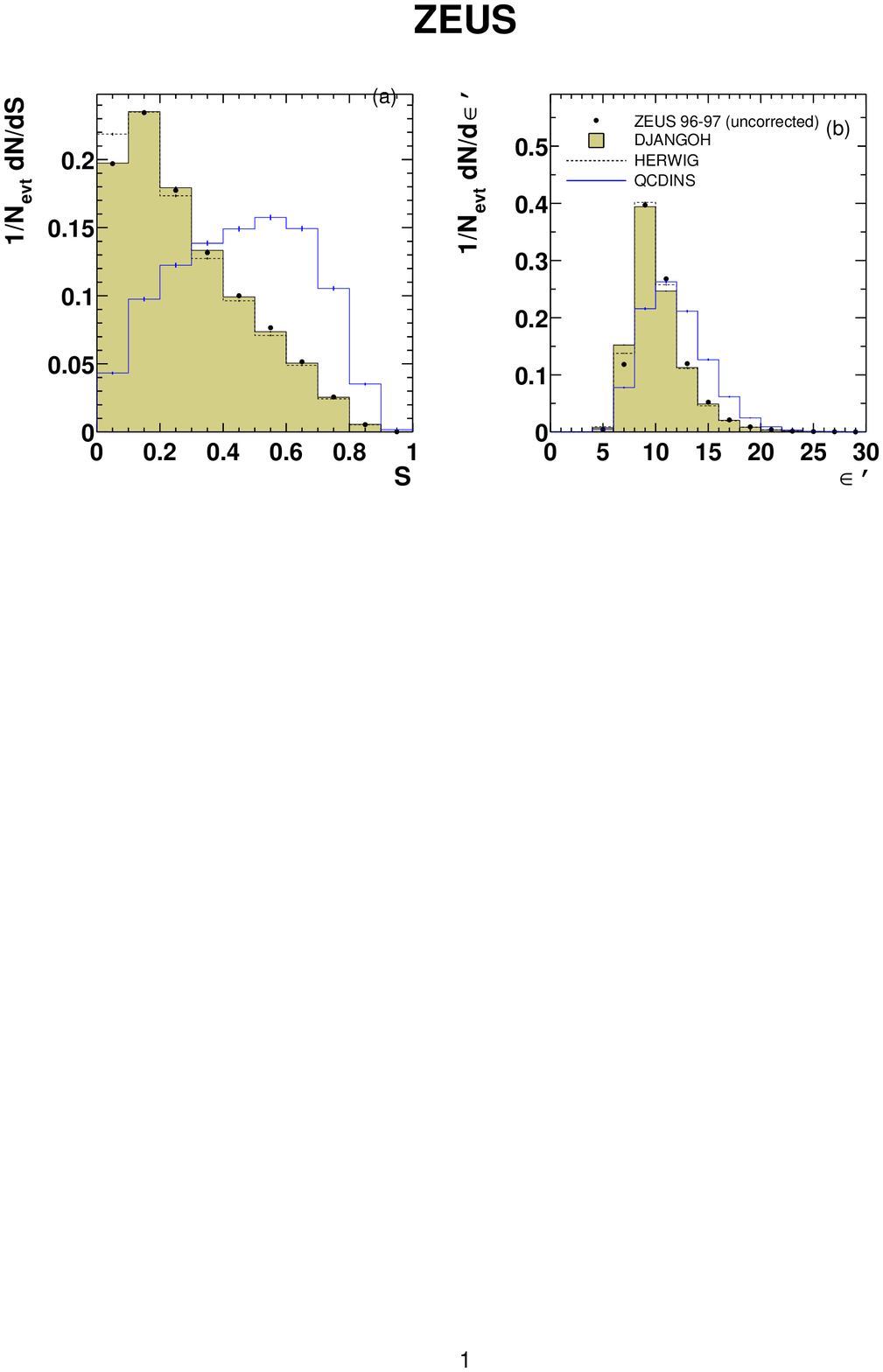,clip=} 
{1.0}
\CondFigEW{FisherDiscriminant}    
{Fisher discriminant $t$ from variables
$S$, $C$, $\mathrm{log_{10}}\,p_T^{\mathrm{Jet}}$, $\NzufosI$,
$\Nzuftr$ and $\epsilon'$}
{Fisher discriminant $t$ calculated from the variables
$S$, $C$, $p_T^{\mathrm{jet}}$, $\NzufosI$,
$\Nzuftr$ and $\epsilon'$.
Shown are (a) linear and (b) logarithmic plots for the fiducial sample
with an additional cut ${Q'}^2_\DA < 250\, \mathrm{GeV}^2$.
In (b), the QCDINS distribution, normalised to the predicted fraction of 
$0.79\%$, is also shown.
}
{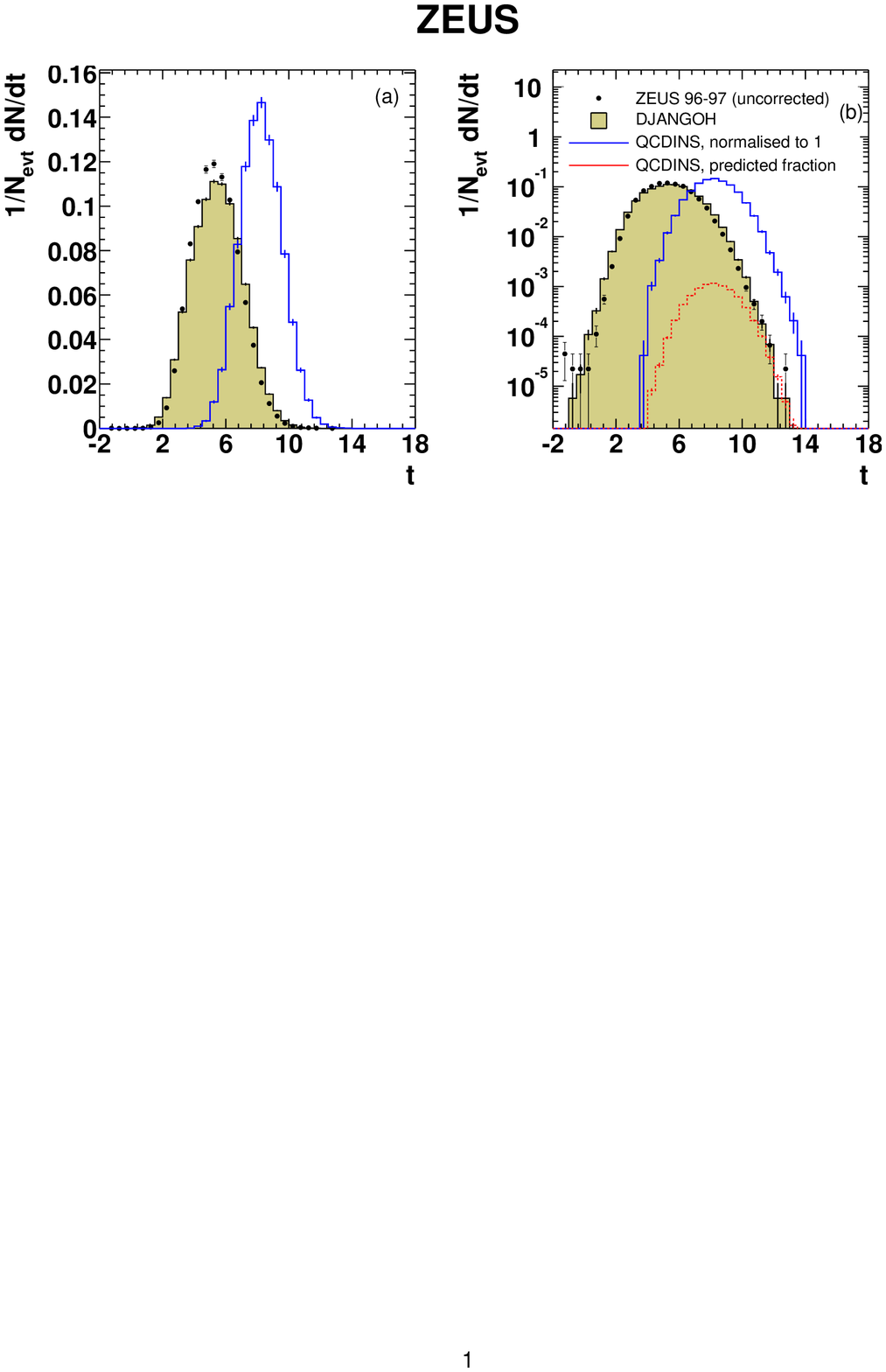,clip=} 
{1.0}
%
%
%
\CondFigEW{Factor}  
{$N_{\mathrm{Data}}/N_{\mathrm{inst}}$ vs $r_I$}
{Plot of the ratio of the cross sections
 $\sigma_d / \sigma_{\rm th}$ vs. $r_I$, the fraction of
instanton-induced events remaining after the specific cut in $t$.
The cross sections $\sigma_d$ and $ \sigma_{\rm th}$ refer
 to the measured cross-section and to the cross-section
 of instanton-induced events according
to the QCDINS Monte Carlo generator, respectively. Inner error bars correspond
to statistical errors, outer error bars show statistical and systematic errors
added in quadrature. The scale on the right hand side gives the
measured cross section in the kinematic region of the fiducial sample.
The dashed line is a conservative 95\% c.l. (see text), corresponding to
$r_I=4.6\%$.}
{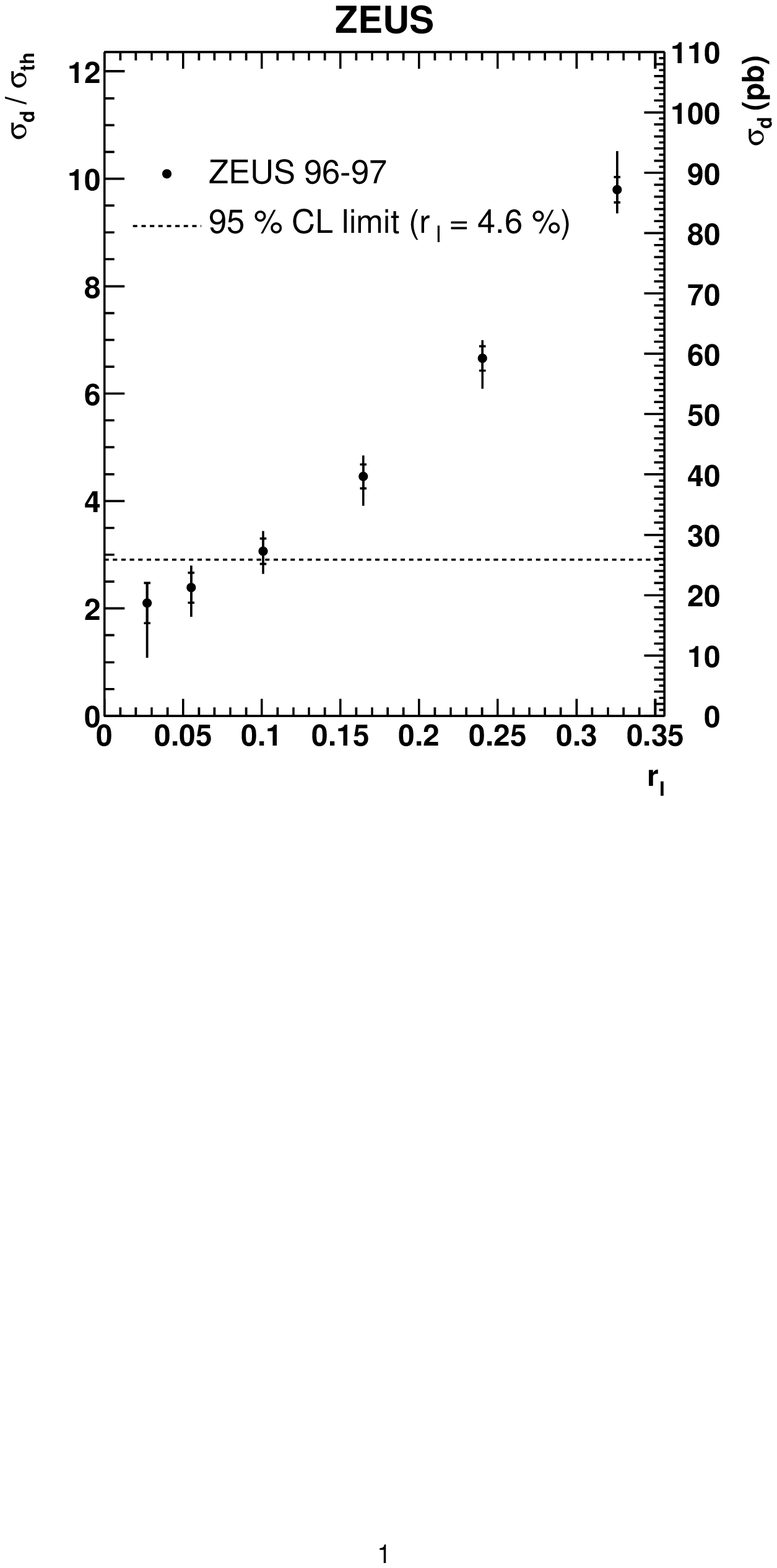,clip=}
{0.65}
%
%
%